\documentclass[trackchanges]{aastex701}

\begin{document}

\title{Statistical Study of Appearance Timing of H$\alpha$ Postflare Loops: Simple Scaling Law Based on Radiative Cooling}

\author[orcid=0009-0004-2275-3991]{Takato Otsu}
\affiliation{Astronomical Observatory, Kyoto University, Sakyo, Kyoto, 606-8502, Japan}
\email[show]{t.otsu@kusastro.kyoto-u.ac.jp}  

\author[orcid=0000-0002-5279-686X]{Ayumi Asai} 
\affiliation{Astronomical Observatory, Kyoto University, Sakyo, Kyoto, 606-8502, Japan}
\email{asai@kwasan.kyoto-u.ac.jp}

\author[orcid=0000-0002-5978-057X]{Kai Ikuta}
\affiliation{Department of Social Data Science, Hitotsubashi University, 2-1 Naka, Kunitachi, Tokyo 186-8601, Japan}
\email{kaiikuta.astron@gmail.com}

\author[orcid=0000-0003-1206-7889]{Kazunari Shibata}
\affiliation{Kwasan Observatory, Kyoto University, Yamashina, Kyoto 607-8471, Japan}
\affiliation{School of Science and Engineering, Doshisha University, Kyotanabe, Kyoto 610-0321, Japan}
\email{shibata@kwasan.kyoto-u.ac.jp}

\begin{abstract}
Recent Sun-as-a-star studies have shown that postflare loops can manifest as a secondary peak in the H$\alpha$ light curve, suggesting that stellar postflare loops are detectable. To understand what determines the timing of such a secondary peak in the H$\alpha$ light curve associated with postflare loops, we must quantitatively identify the key physical processes controlling the appearance of H$\alpha$ postflare loops.
Previous case studies have indicated that the appearance timing of H$\alpha$ postflare loops is likely governed by radiative cooling.
However, the statistical characteristics of the timing of H$\alpha$ postflare loops appearance remain insufficiently investigated.
In this study, we statistically investigated the appearance timing of H$\alpha$ postflare loops to quantify their cooling processes.
As a result, we found a negative correlation between the time difference between the soft X-ray peak and the appearance of the H$\alpha$ postflare loops ($\Delta t$) and the soft X-ray peak flux ($F_\mathrm{X}$).
This relationship is consistent with the theoretical scaling between radiative cooling timescale ($\tau_{\mathrm{rad}}$) and $F_\mathrm{X}$, where $\tau_{\mathrm{rad}} \propto~F_\mathrm{X}^{-1/2}$.
This statistical result indicates that the appearance timing of H$\alpha$ postflare loops relative to the soft X-ray peak is primarily controlled by radiative cooling. Furthermore, we examined the dependence of the scaling law on flare spatial scales ($L$). Consequently, we demonstrated that spatial scale of unresolved stellar flares can be estimated using the following scaling law: $L\propto F_\mathrm{X}^{1/3}\Delta t^{2/3}$. Our results are useful for interpreting secondary peaks in the H$\alpha$ data
of stellar flares and provide new method to estimate spatial scale of
unresolved stellar flares.
\end{abstract}

\section{Introduction} \label{sec:intro}
Postflare loops are loop-shaped plasma structures observed during solar flares, and they are key components of the standard flare model \citep[e.g.,][]{ShibataMagara2011}.
In solar flares, energy released via magnetic reconnection is transported along the magnetic loops toward the solar surface, causing chromospheric evaporation. This process fills the loops with dense and hot plasma that emits soft X-rays (typically $10^7$ K).
Hot loops are thought to cool mainly through radiative cooling, which is enhanced by the high plasma density \citep[e.g.,][]{Aschwanden2001SoPh..204...91A}. Thereafter, the cooler loops are observed in extreme ultraviolet (EUV; typically $10^5-10^6$ K) and chromospheric lines such as H$\alpha$ (typically $10^4$ K).
The cooler loops are particularly called postflare loops, although the term "postflare loops" is sometimes avoided because magnetic reconnection and energy release continue even while cool loops are observed \citep{Svestka2007SoPh..246..393S}.
During the decay phase of flares, downflows of condensed plasma along the postflare loops are sometimes observed \citep[e.g.,][]{Heinzel1992SoPh..139...81H}. 
Their dynamics are observationally investigated using chromospheric and transition region lines \citep[e.g.,][]{Sahin2024ApJ...970..106S,Song2025ApJ...985...52S}.
Although the detailed properties of downflows along postflare loops are still unclear, these downflows are believed to be formed via thermal instability --which can also explain formation of coronal rains and prominences -- triggered by efficient radiative cooling \citep[e.g.,][]{Antolin2022FrASS...920116A}.

Explosive phenomena like solar flares are also observed on various cool stars, and they are called stellar flares.
Stellar flares appear as sudden brightenings on distant stars \citep[e.g.,][]{Kowalski2024LRSP...21....1K}.
In particular, stellar flares that emit energies exceeding $10^{33}$ erg are referred to as superflares \citep[e.g.,][]{MaeharaETAL2012}.
Since such energetic events are thought to significantly affect environments of exoplanets orbiting the host star, superflares are actively investigated through collaborations between the stellar and exoplanet research communities \citep[e.g.,][]{AirapetianETAL2020,CliverETAL2022}.
Moreover, focusing on distant stars allows for collecting more samples of energetic flares. 
Therefore, studies of superflares provide key insights into how large flares can potentially occur on the Sun \citep[e.g.,][]{ShibataETAL2013}.

Recently, a lot of studies have performed spectroscopic observations of stellar flares at optical wavelengths and have reported blue/red shifted emission/absorption in the chromospheric lines such as H$\alpha$ spectra \citep{Vida2019A&A...623A..49V,MaeharaETAL2021, NamekataETAL2022a, InoueETAL2023ApJ,Namekata2024ApJ...961...23N, Kajikiya2025ApJ...979...93K,MuhekiETAL2020,Notsu2024ApJ...961..189N,Leitzinger2024MNRAS.tmp.1385L}. The Doppler-shifted signatures in chromospheric lines imply the existence of cool materials moving on stars.
However, the interpretations of Doppler signatures of stellar flares are basically difficult, since the stellar surfaces cannot be spatially resolved unlike the solar case.
To support the interpretation of stellar spectroscopic data, solar data are utilized via Sun-as-a-star analyses in which solar data are spatially integrated to be directly compared with stellar observations \citep{NamekataETAL2022a,OtsuETAL2022,Ma2024ApJ...966...45M,Pietrow2024A&A...682A..46P,DeWilde2025A&A...700A.275D}.
For example, \cite{NamekataETAL2022a} and \cite{Namekata2024ApJ...961...23N} reported blueshifted absorption and emission signatures, respectively, in the H$\alpha$ spectra of a young solar-type star, EK Draconis.
The temporal evolution of these features closely resembles Sun-as-a-star observations of solar filament/prominence eruptions.  
Therefore, the blueshifted signatures in \citet{NamekataETAL2022a,Namekata2024ApJ...961...23N} are interpreted as stellar filament/prominence eruptions.
These probable stellar filament/prominence eruptions are reproduced by simple pseudo-two-dimensional (magneto-) hydrodynamic simulations of filament/prominence eruptions, which can also explain solar eruptions \citep{Ikuta2024ApJ...963...50I, Namekata2024ApJ...976..255N}, thereby making the above interpretations, especially the existence of associated coronal mass ejections (CMEs), more reliable.

Along with filament/prominence eruptions, postflare loops are important components of the standard flare model.
Therefore, postflare loops are also expected to be observed in stellar cases. Contributions of postflare loops to radiation of stellar flares have been discussed in recent studies \citep[e.g.,][]{Heinzel2018ApJ...859..143H,Yang2023ApJ...959...54Y,Bicz2024ApJ...972L..11B, Odert2025MNRAS.537..537O,Ichihara2025PASJ..tmp...85I}, and downflows along postflare loops have been proposed as possible interpretations of Doppler-shifted signatures in H$\alpha$ stellar spectra \citep{Wu2022ApJ...928..180W,Wollmann2023A&A...669A.118W, NamizakiEtal2023ApJ, Notsu2024ApJ...961..189N, Kajikiya2025ApJ...979...93K}.
However, compared to plasma eruptions, the dynamics of stellar postflare loops remain poorly understood.
Clarifying the dynamics of stellar postflare loops is essential for a comprehensive understanding of stellar flare radiation, particularly during the decay phase.
Recently, \cite{Otsu2024ApJ...974L..13O} performed a Sun-as-a-star analysis of an X1.6 solar flare on 2023 August 5 that exhibited typical dynamics of postflare loops.
As a result, the three important dynamics of postflare loops -- cooling, downflows, successive formation of higher loops -- were confirmed even in the spatially integrated (Sun-as-a-star) data.
Notably, the H$\alpha$ light curve exhibited two peaks: the first peak corresponds to the flare ribbons, occurring almost simultaneously with the soft X-ray peak, and the secondary peak corresponds to the postflare loops. This secondary peak of the H$\alpha$ light curve delayed with respect to the soft X-ray peak, and \citet{Otsu2024ApJ...974L..13O} confirmed that the time difference between them is roughly consistent with the radiative cooling time.
Therefore, simultaneous observations of the soft X-ray peak and the secondary H$\alpha$ peak can serve as a powerful tool for investigating stellar postflare loops through their cooling processes.
Following \citet{Otsu2024ApJ...974L..13O}, \citet{Ichihara2025PASJ..tmp...85I} analyzed a stellar flare on an M dwarf and found a secondary peak in the H$\alpha$ light curve.
Although soft X-ray data were not included in the analysis of the stellar flare presented by \citet{Ichihara2025PASJ..tmp...85I}, the similarity of the H$\alpha$ data to the solar data presented in \citet{Otsu2024ApJ...974L..13O} suggests dynamic behavior of stellar postflare loops.

Although secondary peaks in H$\alpha$ are useful for investigating stellar postflare loops, they can also be caused by other factors, such as prominence eruptions \citep{Inoue2024PASJ...76..175I} or the occurrence of another flare \citep[e.g.,][]{OtsuETAL2022, DeWilde2025A&A...700A.275D}.
To enable a quantitative comparison between solar and stellar postflare loops, a statistical analysis of solar postflare loops is essential.
In particular, to understand what determines the timing of the H$\alpha$ peak associated with postflare loops, we must quantitatively identify the key physical processes controlling the appearance of H$\alpha$ postflare loops.
Although previous studies have suggested that the cooling of flare loops during the decay phase is primarily driven by radiative processes \citep[e.g.,][]{Kamio2003SoPh,Aschwanden2001SoPh..204...91A}, the statistical characteristics of the timing at which H$\alpha$ postflare loops appear remain insufficiently investigated.

In this paper, we present a statistical investigation of the appearance timing of H$\alpha$ postflare loops in solar flares.
This study aims to bridge the gap between solar and stellar flare observations, particularly in interpreting secondary peaks seen in stellar H$\alpha$ light curves.
We determine the time difference ($\Delta t$) between soft X-ray peak and the appearance of H$\alpha$ postflare loops.
We then explore the relationship between $\Delta t$ and the soft X-ray peak flux, providing key insights into the interpretation of secondary peaks in H$\alpha$ light curves.
Since Sun-as-a-star H$\alpha$ light curves are strongly affected by weather conditions such as clouds and atmospheric scintillation, the number of data samples for events with a secondary peak corresponding to postflare loops is insufficient for statistical analysis.
In contrast, the appearance timing of postflare loops can be readily obtained from H$\alpha$ imaging observations even under slightly poor weather conditions.
Therefore, in this study, the appearance timing of H$\alpha$ postflare loops is derived from imaging data to obtain a larger number of data samples.
The observations and analysis methods are presented in Section \ref{sec:obs}.
The results are described in Section \ref{sec:results}, followed by discussions in Section \ref{sec:Discussion} and conclusions in Section \ref{Sec:conc}.

\section{Observations and Analyses} \label{sec:obs}
\begin{figure}[htbp]
\centering
\includegraphics[width=18cm]{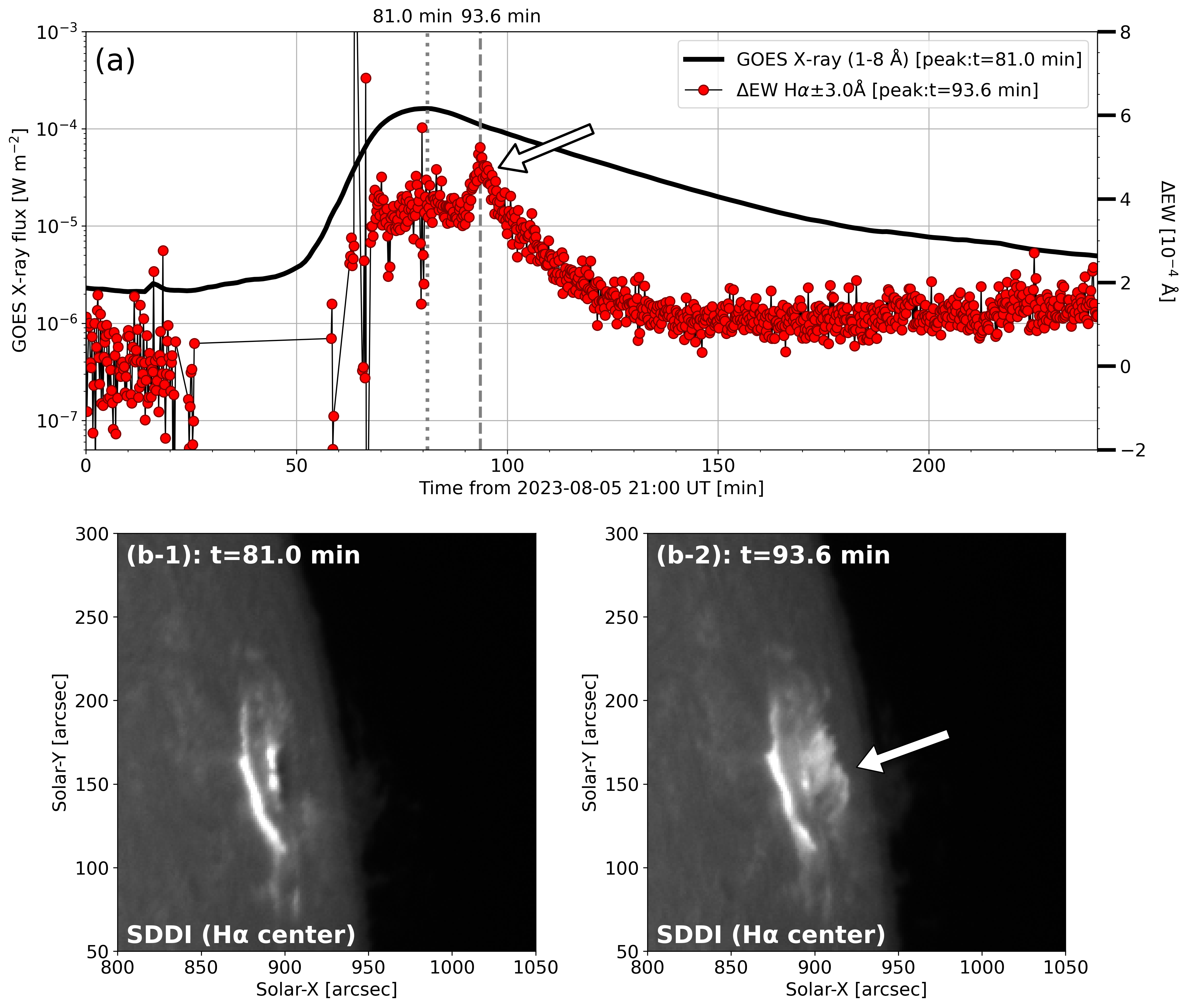}
\caption{(a) The GOES soft X-ray (1–8 {\AA}) light curve is shown as a black solid line. The Sun-as-a-star H$\alpha$ light curve of the X1.6 flare on 2023 August 5 (event 23 in Table \ref{Elist}) is shown as red circles (differenced equivalent width $\Delta$EW of H$\alpha\pm3.0$ {\AA}; see \citet{Otsu2024ApJ...974L..13O} for details). 
The H$\alpha$ data is from \citet{Otsu2024ApJ...974L..13O}.
The white arrow indicates the secondary peak of the H$\alpha$ light curve associated with postflare loops.
The vertical dotted and dashed lines indicate the times of the GOES peak and H$\alpha$ secondary peak, respectively.
(b-1) and (b-2) H$\alpha$ center images taken by SDDI at $t=81$ min (the GOES peak) and 93.6 min (the H$\alpha$ secondary peak) from 21:00 UT on 2023 August 5.
The white arrow in panel (b-2) indicates the postflare loops.}

\label{Otsu2024}
\end{figure}

Figure \ref{Otsu2024} shows the result of the Sun-as-a-star analysis of the X1.6 flare analyzed in \citet{Otsu2024ApJ...974L..13O}, which exhibited the secondary peak corresponding to the postflare loops in the H$\alpha$ light curve. 
In panel (a) of Figure \ref{Otsu2024}, the Sun-as-a-star H$\alpha$ light curve (differenced equivalent width $\Delta$EW of H$\alpha\pm3.0$ {\AA}; see \citet{Otsu2024ApJ...974L..13O} for details) is shown together with the GOES soft X-ray light curve.
The white arrow in panel (a) of Figure \ref{Otsu2024} indicates the secondary peak of the Sun-as-a-star H$\alpha$ light curve.
At the GOES soft X-ray peak, two ribbons can be seen in the H$\alpha$ image (Figure \ref{Otsu2024} (b-1)).
In contrast, corresponding to the secondary peak of the Sun-as-a-star H$\alpha$ light curve, the postflare loops appear as bright features in the H$\alpha$ center image as indicated by the white arrow in Figure \ref{Otsu2024} (b-2).
Although this study is motivated by such secondary peak in the Sun-as-a-star H$\alpha$ light curve, we derived the appearance timing of H$\alpha$ postflare loops from imaging data in order to obtain a larger number of data samples. The effect of this methodological difference on the results will be discussed in Section \ref{sub:ToSaS}.

\subsection{Instruments and examples of H$\alpha$ observations}
We analyzed 34 solar flares associated with postflare loops observed in H$\alpha$ images.
To determine the appearance times of H$\alpha$ postflare loops ($t_{\mathrm{H\alpha}}$), we used H$\alpha$ images taken by four different instruments. In the following, we introduce these four instruments and present examples of H$\alpha$ images obtained from each instrument.

\subsubsection{The Sartorius 18 cm refractor telescope (Sartorius)}
\begin{figure}[htbp]
\centering
\includegraphics[width=11.5cm]{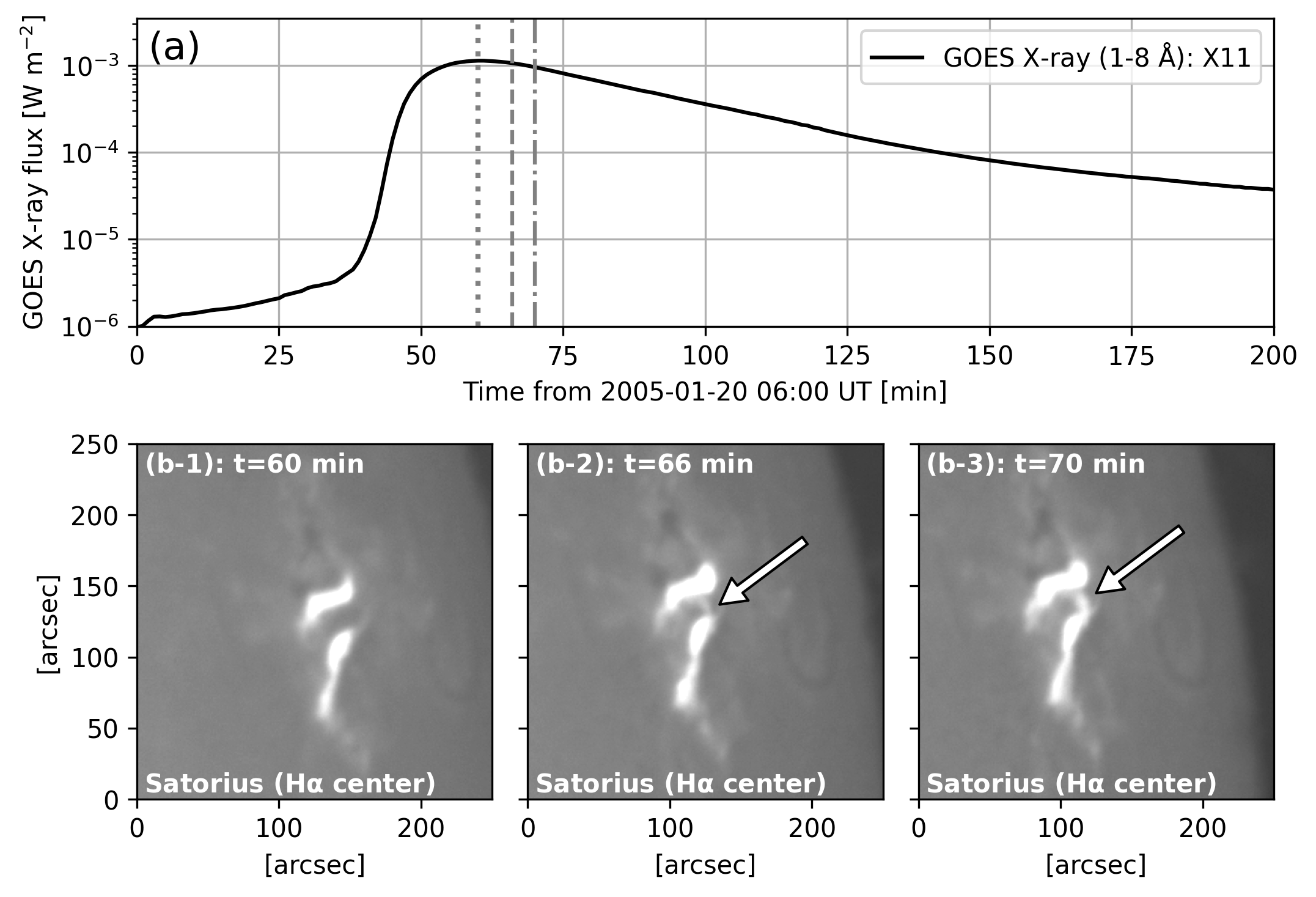}
\caption{Overview of the X11 flare on 2005 January 20 (event 3). (a)The light curve of GOES soft X-ray (1-8 $\mathrm{\AA}$) flux is shown as the black solid line.
The vertical gray dotted, dashed, and dotted-dashed lines indicate the times of the GOES peak, H$\alpha$ loops appearance, and panel (b-3), respectively. (b-1)–(b-3) H$\alpha$ center images taken by Sartorius are shown at the times of 60, 66, and 70 minutes from 06:00 UT on 2005 January 20. The times of panels (b-1) and (b-2) correspond to the GOES peak and the appearance of H$\alpha$ loops, respectively. The white arrows in (b-2) and (b-3) indicate the postflare loops.}
\label{18cm01}
\end{figure}

The \textit{Sartorius} 18 cm refractor telescope \citep{Asai2003ApJ...586..624A} is located at Kwasan Observatory, Kyoto University.
\textit{Sartorius} observes monochromatic H$\alpha$ images of the Sun using a Halle Lyot filter with a 0.5 {\AA} bandwidth in the H$\alpha$ line center. The pixel size and the time cadence are about 1$^{\prime\prime}$.06 and 1 second, respectively.
Figure \ref{18cm01} shows the temporal evolution of the X11 flare that occurred on 2005 January 20, as observed by \textit{Sartorius} (event 3 in Table \ref{Elist}). Figure \ref{18cm01} (a) displays the GOES soft X-ray light curve of this flare. The bottom panels, from (b-1) to (b-3), show H$\alpha$ center images taken by \textit{Sartorius} at the GOES peak time, at the appearance time of the H$\alpha$ postflare loops, and at the time when H$\alpha$ postflare loops were clearly visible, respectively.
The white arrows in panels (b-2) and (b-3) indicate the postflare loops. 

\subsubsection{The Solar Magnetic Activity Research Telescope/T1 (SMART/T1)}\label{T1}
T1 installed on the Solar Magnetic Activity Research Telescope \citep[SMART;][]{UenoETAL2004} at Hida Observatory, Kyoto University regularly observed the full-disk Sun at multiple wavelengths: H$\alpha$ line center, H$\alpha\pm0.5$ {\AA}, H$\alpha\pm0.8$ {\AA}, and H$\alpha\pm1.2$ {\AA}, with the time cadence of 1 to 2 minutes and the pixel size of 0$^{\prime\prime}$.56.
SMART/T1 has now been replaced by SMART/SDDI (see Section \ref{inst:SDDI}).
Figure \ref{smart01} is the same format as Figure \ref{18cm01}, but for the X7.8 flare that occurred on 2012 March 7 and was observed by SMART/T1 (event 6 in Table \ref{Elist}).
The white arrows in panels (b-2) and (b-3) indicate the postflare loops. 
This flare was previously analyzed in detail by \cite{Takahashi2015ApJ...801...37T} in the context of the prominence oscillation (activation) driven by the associated EUV wave.

\subsubsection{Flare Imaging System in Continuum and H$\alpha$ (FISCH)}
The Flare Imaging System in Continuum and H$\alpha$ \citep[FISCH;][]{Ishii2013PASJ...65...39I} was installed on SMART in 2011 August.
FISCH has a field of view of 344$^{\prime\prime} \times$ 258$^{\prime\prime}$, the pixel size of 0$^{\prime\prime}$.214, and the time cadence of 30–33 frames per second.
Figure \ref{FISCH01} is in the same format as Figure \ref{18cm01}, but for the X1.9 flare that occurred on 2013 May 15 and was observed by FISCH (event 10 in Table \ref{Elist}).
The white arrows in panels (b-2) and (b-3) indicate the postflare loops. 

\subsubsection{The Solar Dynamics Doppler Imager (SDDI)}\label{inst:SDDI}
The Solar Dynamics Doppler Imager \citep[SDDI;][]{IchimotoETAL2017} was installed on SMART/T1 in 2016 May.
SDDI regularly takes full-disk images of the Sun at wavelengths ranging from H$\alpha-9.0$ {\AA} to H$\alpha+9.0$ {\AA}, with a spectral resolution of 0.25 {\AA}.
The time cadence and the pixel size are 12 seconds and 1$^{\prime\prime}$.23, respectively.
Figure \ref{sddi01} is in the same format as Figure \ref{18cm01}, but for the M1.1 flare that occurred on 2021 April 19 and was observed by SDDI (event 13 in Table \ref{Elist}).
In panels (c-1) to (c-3), H$\alpha+0.75$ {\AA} images are displayed.
Although these events did not show clear postflare loops in H$\alpha$ center images (Figure \ref{sddi01} (b-2)-(b-3)), dark downflows can be seen in H$\alpha$ red-wing images (Figure \ref{sddi01} (c-2)-(c-3)) 
This event was analyzed by \citet{OtsuETAL2022} in the context of the Sun-as-a-star study.

\subsection{Event List}\label{Event}
We analyzed 34 flares that exhibited postflare loops in H$\alpha$ images observed by the above instruments. For SDDI data, we obtained events from the event list in 2023 and 2024.
We also used events from 2017 to 2022 (events 12–15), some of which were analyzed in previous studies \citep{OtsuETAL2022, Otsu2024ApJ...964...75O}.
As for the SMART/T1, FISCH, and \textit{Sartorius}, we used only major events captured by them.
Finally, we selected 3, 6, 3, and 22 events from \textit{Sartorius}, SMART/T1, FISCH, and SDDI, respectively.
We summarized the GOES peak time ($t_\mathrm{X}$), the appearance time of H$\alpha$ postflare loops ($t_\mathrm{H\alpha}$), time difference ($\Delta t$), GOES peak flux ($F_\mathrm{X}$), volume emission measure ($EM$), temperature ($T$), ribbon distance ($d$), square root of flare area ($A^{1/2}$), flare spatial scale ($L$), instrument for H$\alpha$ observation, event type, and NOAA active region number in Table \ref{Elist}.
Detailed definitions of event type and parameters ($t_\mathrm{X}$, $t_\mathrm{H\alpha}$, $\Delta t$, $F_\mathrm{X}$, $EM$, $T$, $d$, $A$, $L$) are described in the next subsection.

\subsection{Analyses}
\subsubsection{Event type}\label{Etype}
The event list includes both off-limb events and on-disk events.
The off-limb events are labeled as ``off''in Table \ref{Elist}.
On-disk events are further classified into two categories based on the visibility of postflare loops in the H$\alpha$ line center images.
On-disk events that exhibit well-defined loop structures connecting two ribbons in the H$\alpha$ line center are classified as ``c'', which stands for ``clear''.
In contrast, on-disk events without clear postflare loops in H$\alpha$ center images are categorized as ``u'', which stands for ``unclear''.
In observations with SDDI, many on-disk events show dark features in the H$\alpha$ red wing, which can be interpreted as downflows along postflare loops (i.e., flare-driven coronal rain). However, some of these events do not exhibit clearly identifiable loop structures in the H$\alpha$ line center images and they are classified as ``u''-type events.
Figure \ref{sddi01} presents a representative example of ``u''-type events. 
Although the loops are not clearly visible in the H$\alpha$ line center (Figure \ref{sddi01} (b-2) and (b-3)), the dark downflows are clearly observed in the H$\alpha+0.75$ {\AA} images (Figure \ref{sddi01} (c-2) and (c-3)).
To determine the appearance timing of H$\alpha$ postflare loops,
we used H$\alpha$ center images for events labeled as ``c'' and ``off'', and H$\alpha$ red wing images (i.e., downflow signatures) for events of type ``u''.
The event types are listed in Table \ref{Elist}.

\subsubsection{Measurement of Time Difference: $\Delta t$}
To investigate what determines the appearance timing of the H$\alpha$ postflare loops, we estimated the time difference between the peak time of GOES soft X-ray flux and the appearance time of H$\alpha$ postflare loops.
As for the appearance timing of the H$\alpha$ loops,
we checked by eyes the time series of H$\alpha$ images for each event and defined the moment when the postflare loops first appear in the H$\alpha$ image as the appearance time of the H$\alpha$ loops ($t_{\mathrm{H}\alpha}$) (Figure \ref{18cm01}, \ref{smart01}, \ref{FISCH01}, and \ref{sddi01}). 
We used H$\alpha$ center images for events labeled as ``c'' and ``off'', and H$\alpha$ red wing images for events of type ``u'', as described in Section \ref{Etype}.
Using $t_{\mathrm{H}\alpha}$ and the GOES peak time ($t_\mathrm{X}$), we calculated the time difference $\Delta t = t_{\mathrm{H}\alpha} - t_\mathrm{X}$.
In some cases (events 22, 33, and 34 in Table \ref{Elist}), H$\alpha$ postflare loops appear earlier than the GOES peak time, although they are typically faint.
In such cases, we defined the moment when the prominent H$\alpha$ loops appeared as $t_{\mathrm{H}\alpha}$. 

\subsubsection{Peak Flux and Emission Measure of GOES soft X-rays}\label{M:GOES}
We used the peak flux of GOES/XRS-B (1-8 {\AA}) ($F_{\mathrm{X}}$) as the representative value of the flare.
It is worth noting that NOAA recalibrated the GOES/XRS data in 2020 May and we used these recalibrated data. For example, event 6 was well known as the ``X5.4'' flare on 2012 March 7, but we used the recalibrated flux value $7.8\times10^{-4}$ W m$^{-2}$ corresponding to X7.8 class as the peak flux of this event (Section \ref{T1}). 
For details, we refer the reader to section 2.2 in the XRS User's Guide: \url{https://data.ngdc.noaa.gov/platforms/solar-space-observing-satellites/goes/goes16/l2/docs/GOES-R_XRS_L2_Data_Users_Guide.pdf}. 
We also calculated the volume emission measure ($EM$ [cm$^{-3}$]; hereafter emission measure) and temperature ($T$) from the GOES/XRS-A (0.5-4 {\AA}) and -B fluxes at the time of the GOES/XRS-B peak.
We used the Python package {\it sunkit\_instruments} 0.5.0 \citep{Sunpy2020ApJ} and applied  {\it goes\_xrs.calculate\_temperature\_em} to GOES/XRS-A and B data.
This calculation is based on the method described in \cite{White2005SoPh..227..231W}.
For the input abundance of this calculation, we used the photospheric value, assuming the evaporated plasma is dominated by the photospheric abundance. The GOES/XRS-A channel was used exclusively for the $EM$ calculation. Therefore, for simplicity, we refer to the GOES/XRS-B flux as the GOES flux hereafter.
The obtained $F_{\mathrm{X}}$, $EM$, and $T$ are also listed in Table \ref{Elist}.

\subsubsection{Spatial Scale}\label{M:spatial}
To investigate the details of the relationship between $\Delta t$ and $F_\mathrm{X}$ (or $EM$), we derived the spatial scale $L$ of flares.
In the following method, we used the H$\alpha$ center image with the field of view covering the target flare at the GOES peak time for each event.
Examples of the field of view for events 3, 6, 10, and 13 are shown in Figure \ref{Area}.
First, we obtained the ribbon area $A$ as follows.
We defined the masked region as the area where the H$\alpha$ center intensity exceeds 65\% of its maximum value within the field of view.
Examples of the masked regions for events 3, 6, 10, and 13 are shown as areas enclosed by red solid lines in Figure \ref{Area}.
We obtained the flare area ($A$) by summing the pixels within the masked region.
Second, we obtained the ribbon distance $d$ as the distance between the intensity-weighted centroids of two flare ribbons.
Examples of two centroids and the distances between them for events 3, 6, 10, and 13 are shown as black points and a black dashed line, respectively, in Figure \ref{Area}.
We corrected for the projection effect based on the distance from the solar disk center for both $d$ and $A$.
In Table \ref{Elist}, values of $d$ and $A^{1/2}$ are summarized.
Third, we calculated the flare volume $V$ using $d$ and $A$ under the assumption of the semi-circular geometry.
$A/2$ roughly corresponds to the total cross section of the overall flare loops (arcades), whereas $\pi d/2$ corresponds to the loop length.
By using these two quantities, we calculated the flare volume as $V=(\pi d/2)\times(A/2)$.
Finally, we derived the spatial scale $L$ as the cubic root of $V$: $L=V^{1/3}$.
For the off-limb events (event 4, 8, 17, 24), we could not obtain $d$ and $A$, and simply defined the spatial scale $L$ of these events as the square root of loop area at $t=t_\mathrm{H\alpha}$ using the same method for the derivation of ribbon areas.
Figure \ref{Area} (e-1) and (e-2) show schematic pictures of the estimation of flare spatial scale $L$ for on-disk cases (types ``c'' and ``u'') and off-limb cases (types of ``off''), respectively.

\section{Results} \label{sec:results}
\subsection{Distribution of Parameters}\label{Re:params}
Figure \ref{hist} (a)-(d) presents the histograms of the time difference $\Delta t$, the spatial scale $L$, the temperature $T$, and the emission measure $EM$.
The values of $EM$ and $T$ are measured at the GOES peak times (see Section \ref{M:GOES}).
The time differences range from  1 to 33 minutes, with a mean of approximately 9.5 minutes (Figure \ref{hist} (a)). In this study, we used 1-minute cadence data for H$\alpha$ in accordance with GOES/XRS data.
Consequently, the lower limit of $\Delta t$ is 1 minute; however, only two events fall into this category, so the impact of this limitation on the statistical results is likely negligible.
The spatial scale $L$ ranges from 5.1 to 50.4 Mm, with an average of approximately 19.8 Mm (Figure \ref{hist} (b)), which is roughly consistent with typical spatial scales of solar flares (10-100 Mm) presented in previous studies \citep[e.g.,][]{Shibata1999ApJ...526L..49S,Namekata2017ApJ...851...91N}.
The temperature $T$ ranges from $0.55\times10^7$ to $2.58\times10^7$ K (Figure \ref{hist} (c)), while the emission measure $EM$ ranges from $0.09\times10^{50}$ to $4.85\times10^{50}$ cm$^{-3}$ (Figure \ref{hist} (d)).
The average values of $T$ and $EM$ are approximately $1.3\times10^7$ K and $1.28\times10^{50}$ cm$^{-3}$, respectively. 

\subsection{Relation between $\Delta t$ and $F_{\mathrm{X}}$ ($EM$)}
Figure \ref{types} (a) shows the relation between $\Delta t$ and the GOES peak flux $F_\mathrm{X}$, while Figure \ref{types} (b) presents the relationship between $\Delta t$ and $EM$. 
Notably, both relationships exhibit clear anti-correlations. 
Correlation coefficients for $F_\mathrm{X}$-$\Delta t$ and $EM-\Delta t$ relationships on the log-log scale are $-0.64$ and $-0.61$, respectively.
We performed a linear fit to the data points in log–log space, and obtained $\log_{10}(\Delta t)=(-0.39\pm0.03)\log_{10}(F_\mathrm{X})+(-0.81\pm0.12)$, and $\log_{10}(\Delta t)=(-0.49\pm0.04)\log_{10}(EM)+(25.0\pm2.0)$.
In Figure \ref{types} (a) and (b), the data points are categorized according to the event types described in section \ref{Etype} (listed in Table \ref{Elist}): on-disk events with clear postflare loops in H$\alpha$ center images (type ``c'')  are marked with orange circles; on-disk events without clear postflare loops in H$\alpha$ center images (type ``u'') are marked with black stars; and events showing off-limb postflare loops (type ``off'') are marked with sky-blue triangles.
Events with clear postflare loops and those without clear postflare loops in H$\alpha$ center images are roughly separated at $F_{\mathrm{X}}\approx8\times10^{-5}$ W m$^{-2}$ ( $EM\approx8\times10^{49}$ cm$^{-3}$), which are shown as gray vertical dotted lines in Figure \ref{types} (a) and (b).

Figure \ref{AreaColor} shows the same relationships $F_\mathrm{X}$-$\Delta t$ and $EM$-$\Delta t$ as Figure \ref{types}, but with data points colored according to the spatial scale $L$. 
The relationship between $F_\mathrm{X}$ and $\Delta t$ demonstrates a dependence on $L$: events with smaller (larger) $L$ tend to exhibit shorter (longer) $\Delta t$ at a given $F_\mathrm{X}$.
A similar dependence on $L$ is also evident in the $EM$-$\Delta t$ relationship.
We discuss the comparison of these results with the radiative cooling timescale in Section \ref{Dis:scaling}.

\section{Discussion}\label{sec:Discussion}

\subsection{Flares with and without clearly identifiable postflare loops in H$\alpha$ center images}

In this study, we categorized the events as follows: events with clear postflare loops in H$\alpha$ center (type ``c''); events without clear postflare loops in the H$\alpha$ center images (type ``u''); and events showing off-limb postflare loops (type ``off'').
In Figure \ref{types}, types ``c'' and ``u'' are roughly separated at $F_{\mathrm{X}}\approx8\times10^{-5}$ W m$^{-2}$ ($EM\approx8\times10^{49}$ cm$^{-3}$).
This separation can be explained by the dependence of $F_\mathrm{X}$ ($EM$) on electron density. 
It is known that postflare loops with relatively low density appear as absorption features in H$\alpha$ (e.g., an electron density of around $10^{11}$ cm$^{-3}$ was obtained in \citet{Heinzel1992SoPh..139...81H} for postflare loops observed as absorption in the H$\alpha$ line), whereas only dense loops can be seen as emission in the H$\alpha$ center on the solar disk \citep{Svestka1976GAM.....8.....S}.
This suggests that our events with bright loops in H$\alpha$ center are associated with higher electron densities, which is consistent with the larger $F_\mathrm{X}$ ($EM$) values.
The electron density of our events is to be discussed in Section \ref{Dis:scaling}.
We note, however, that $F_\mathrm{X}$ ($EM$) also depends on the spatial scale, which may account for the ambiguity in the separation between types ``c'' and ``u''.
The brightness of postflare loops in H$\alpha$ is a key factor in determining their contribution to spatially integrated observations \citep{Otsu2024ApJ...974L..13O}. Although the classification of events with or without clear postflare loops was done visually in this study, 
the visibility of postflare loops in H$\alpha$ images and in Sun-as-a-star data needs to be quantitatively investigated. 

\subsection{Relation between H$\alpha$ loops appearance timing and emission measure and soft X-ray peak flux}\label{Dis:scaling}

\begin{figure}[htbp]
\centering
\includegraphics[width=11.5cm]{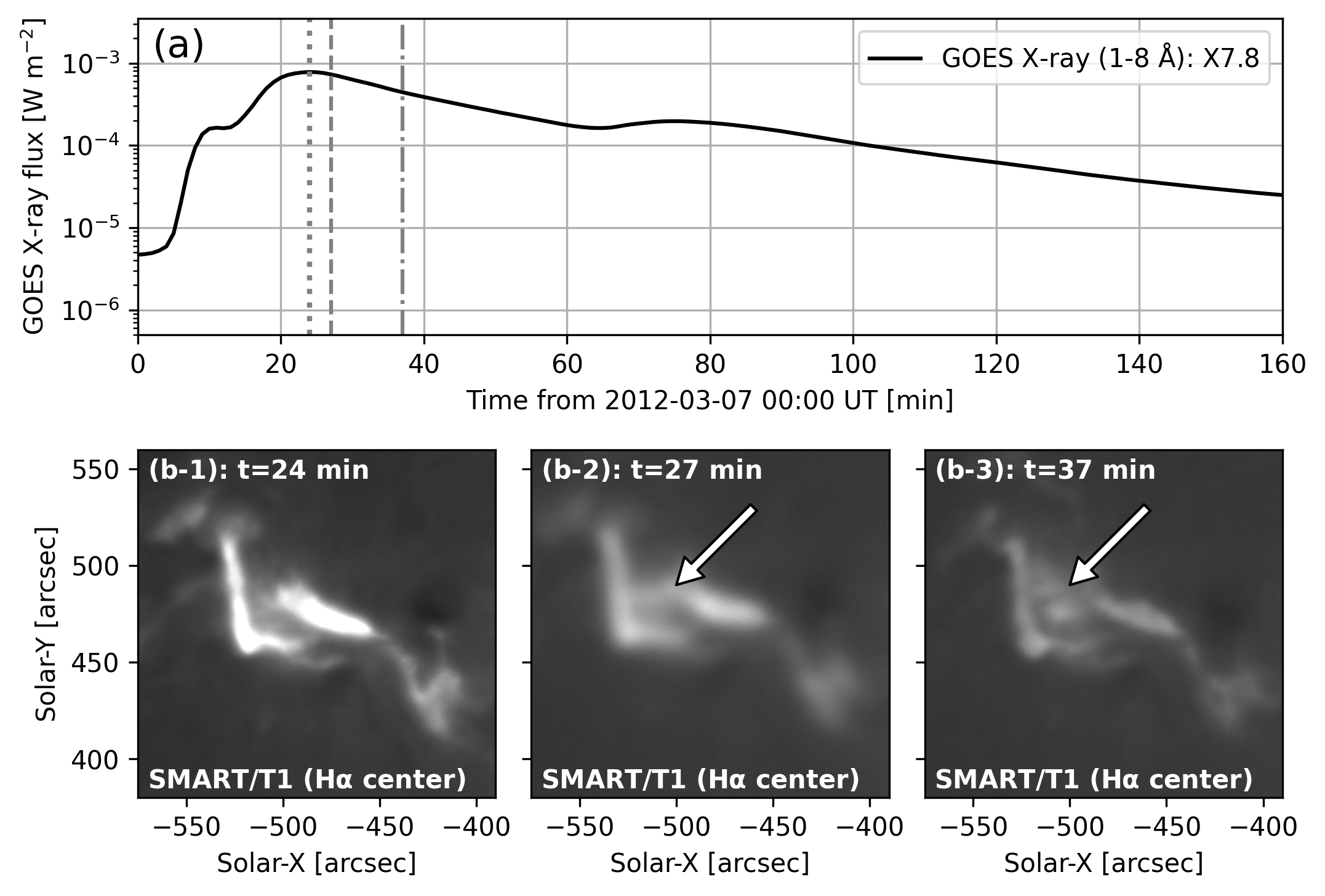}
\caption{Same as Figure \ref{18cm01}, but for the X7.8 flare on 2012 March 7, observed by SMART/T1 (event 6).}
\label{smart01}
\end{figure}

\begin{figure}[htbp]
\centering
\includegraphics[width=11.5cm]{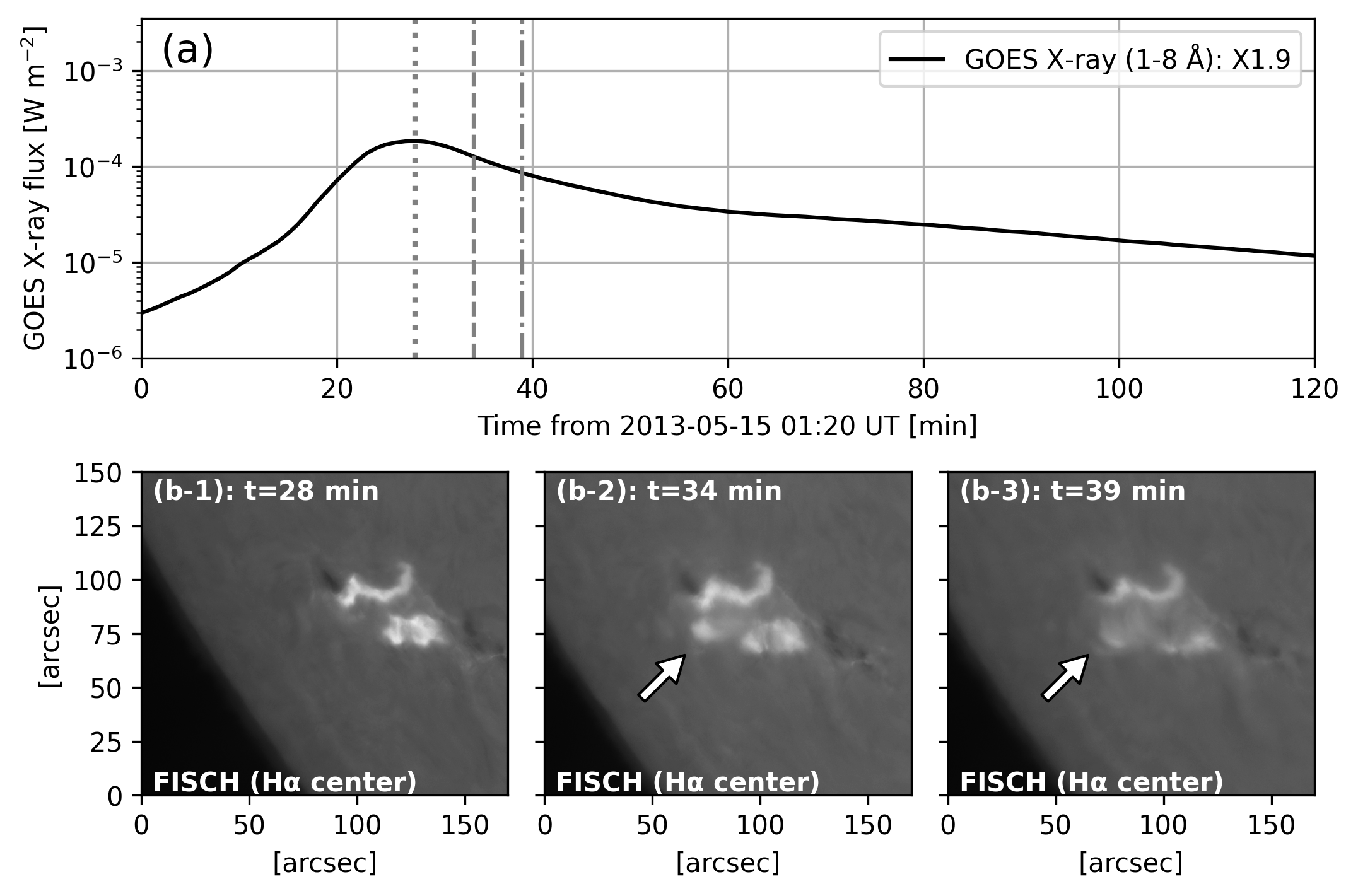}
\caption{Same as Figure \ref{18cm01}, but for the X1.9 flare on 2013 May 15, observed by FISCH (event 10).}
\label{FISCH01}
\end{figure}

\begin{figure}[htbp]
\centering
\includegraphics[width=11.5cm]{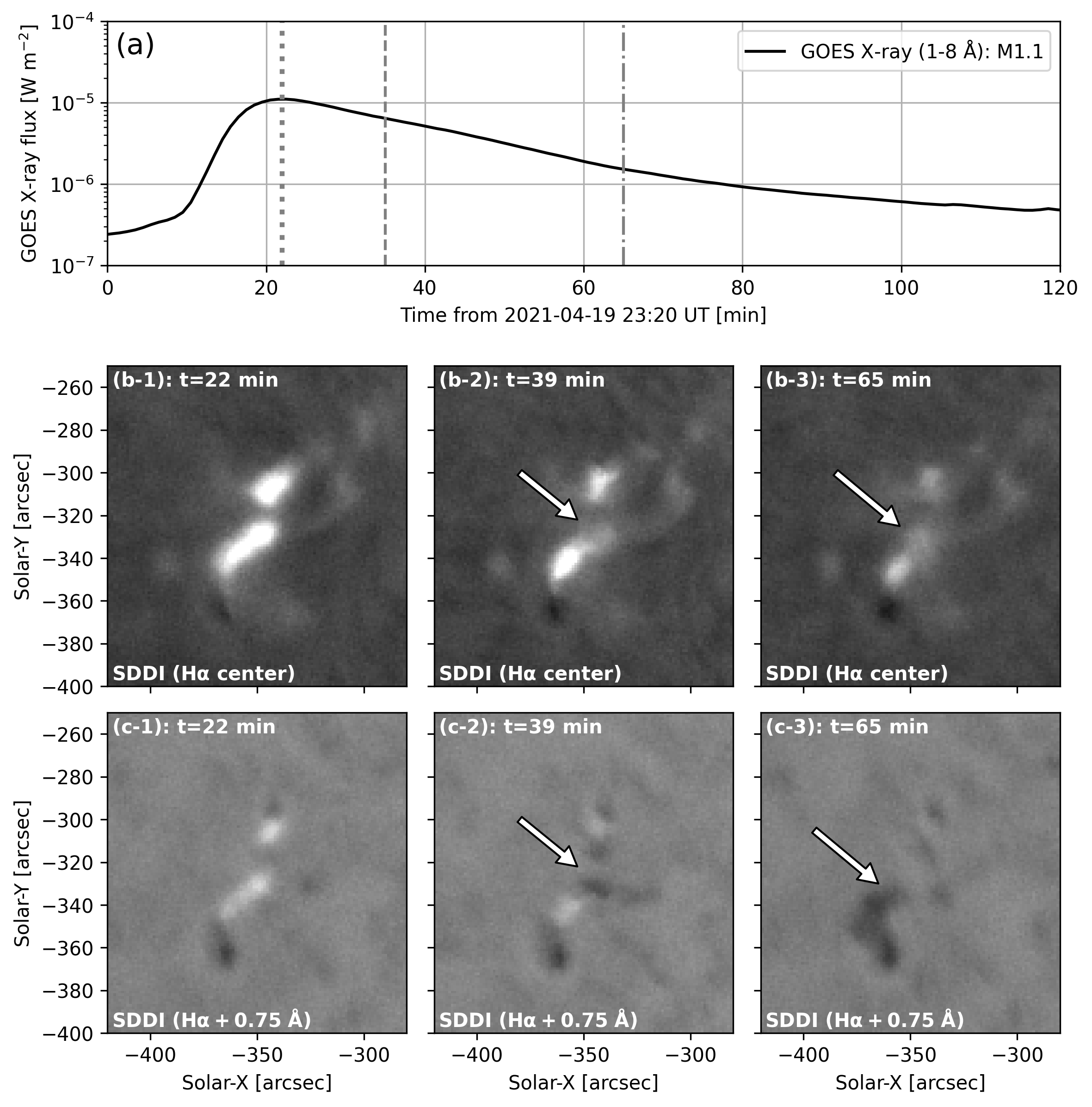}
\caption{Same as Figure \ref{18cm01}, but for the M1.1 flare on 2021 April 19, observed by SDDI (event 13). Panels (c-1)–(c-3) show H$\alpha + 0.75~\mathrm{\AA}$ images in the same format as panels (b-1)–(b-3).
The white arrows in panels (c-2) and (c-3) indicate dark downflows associated with the postflare loops.}
\label{sddi01}
\end{figure}

\begin{figure}[htbp]
\centering
\includegraphics[width=15cm]{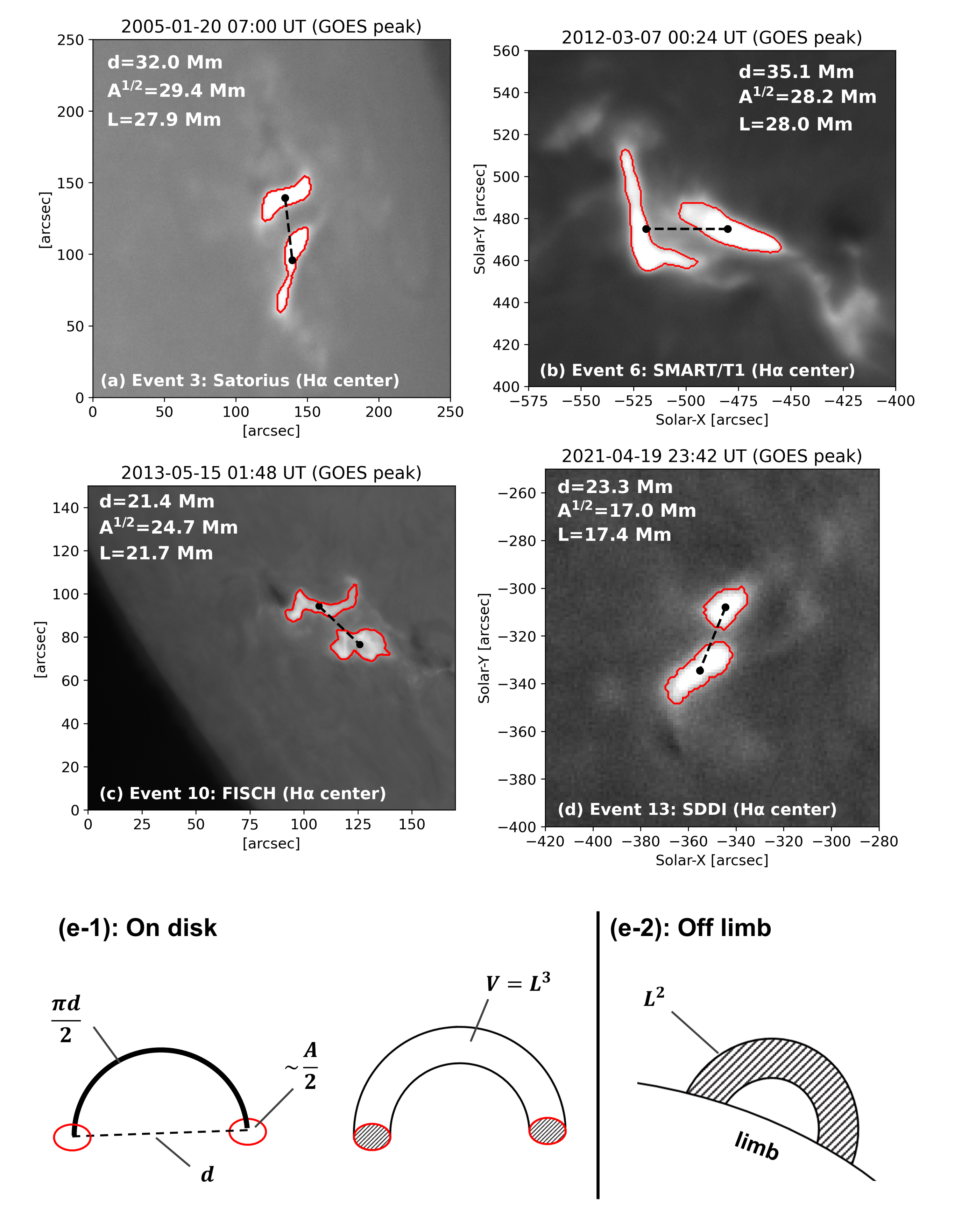}
\caption{Examples and schematic pictures of the estimation of spatial scales. (a)-(d) show the H$\alpha$ center images at the GOES peak time for events 3, 6, 10,and 13, respectively. In each panel, red contours indicate the ribbon regions. The black dashed line connecting the two black points corresponds to ribbon distance. The obtained ribbon distance ($d$), the square root of ribbon area ($A^{1/2}$), and the spatial scales are written in each panel. For the details of these scales, see the text.
(e-1) Schematic picture of the assumed semi-circular arcade for on-disk cases (the types ``c'' and ``u''). $d$, $A$ and $V$ are the ribbon distance, ribbon area, and flare volume, respectively. The cubic root of the flare volume corresponds to the spatial scale $L$. (e-2) Schematic picture  for off-limb cases (the type of ``off''). The dashed region indicates the loop area. The square root of the loop area corresponds to the spatial scale $L$.
}
\label{Area}
\end{figure}

\begin{figure}[htbp]
\centering
\includegraphics[width=15cm]{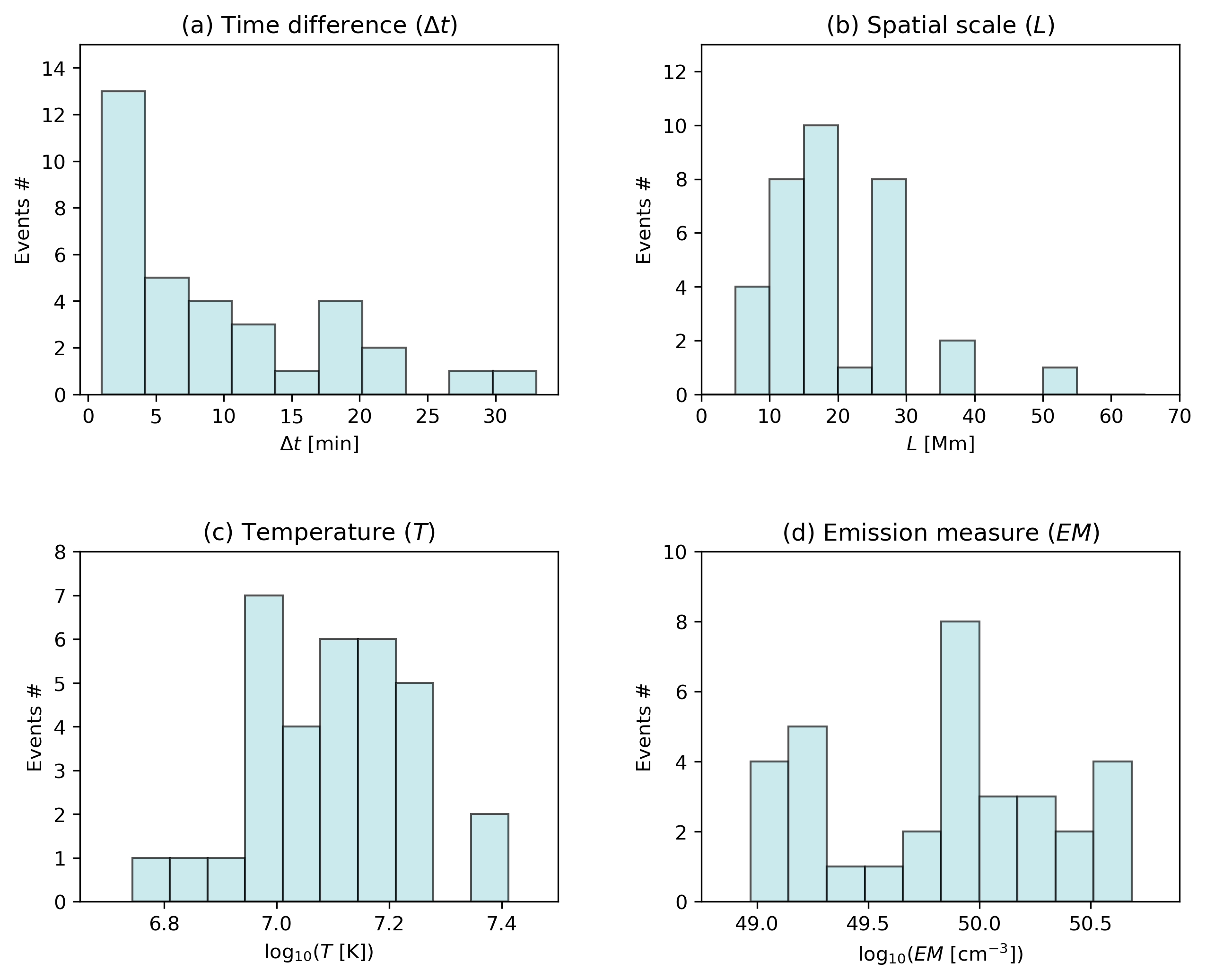}
\caption{Histograms of parameters. (a) Time difference between GOES soft X-ray (1-8 {\AA}) peak time and the appearance time of H$\alpha$ postflare loops ($\Delta t$). (b) Spatial scale ($L$). (c) Temperature ($T$). (d) Emission measure ($EM$). We note that panels (a) and (b) include all 34 events, while panels (c) and (d) include 33 events, excluding event 27.}
\label{hist}
\end{figure}

\begin{figure}[htbp]
\centering
\includegraphics[width=18cm]{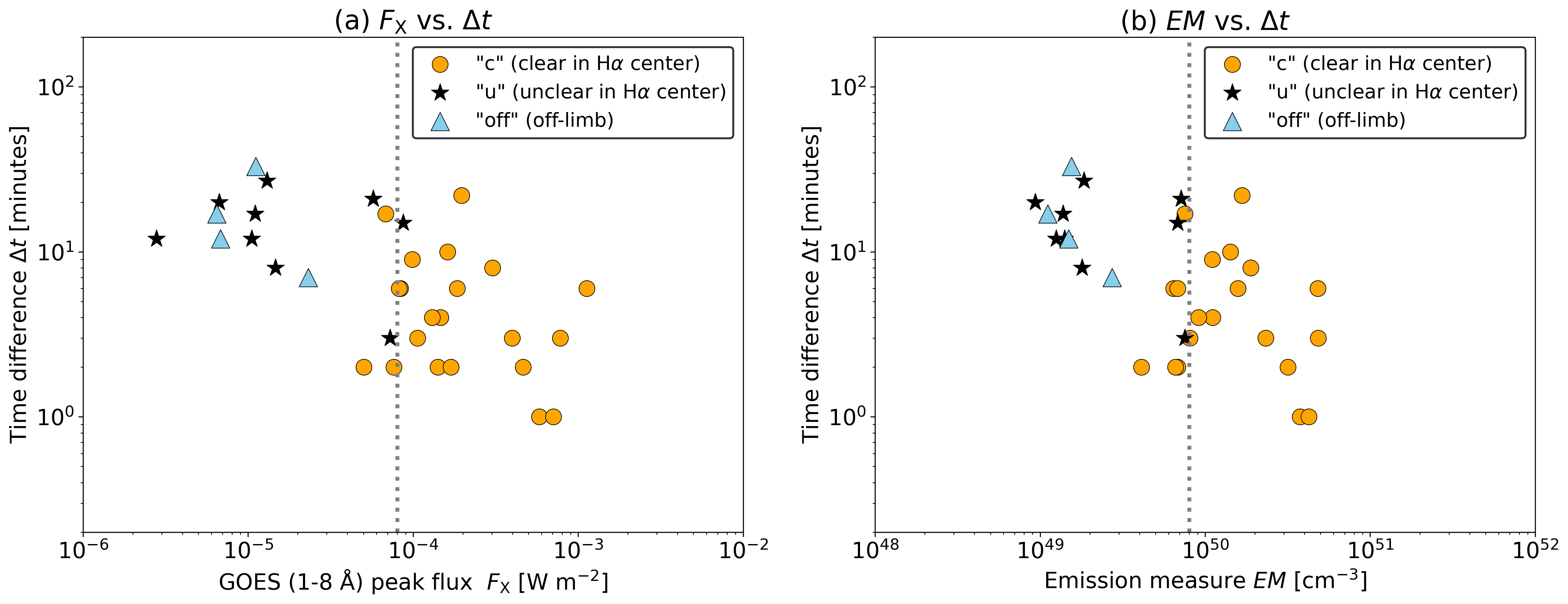}
\caption{(a)The relationship of GOES soft X-ray (1-8 {\AA}) peak flux ($F_\mathrm{X}$) and time difference between GOES soft X-ray (1-8 {\AA}) peak time and the appearance time of H$\alpha$ postflare loops ($\Delta t$). (b)The relationship of emission measure (EM) and $\Delta t$. In both panels, the orange circles indicate events which showed postflare loops clearly in H$\alpha$ center images, whereas the black stars indicate the events without clear postflare loops in H$\alpha$ center images. The skyblue triangles indicate the events with off-limb postflare loops.
The vertical gray dotted lines in panels (a) and (b) indicate $F_\mathrm{X}=8\times10^{-5}$ W $\mathrm{m}^{-2}$ and $EM=8\times10^{49}$ cm$^{-3}$, respectively.
We note that (a) includes all of the 34 events, but (b) includes the 33 events excluding event 27.}
\label{types}
\end{figure}

As shown in Figures \ref{types} and \ref{AreaColor}, $F_\mathrm{X}$-$\Delta t$ and $EM$-$\Delta t$ exhibit negative correlations.
Here, we discuss an explanation for the relation between $F_\mathrm{X}$ ($EM$) and $\Delta t$ based on radiative cooling.
We obtain the radiative cooling time scale $\tau_\mathrm{rad}$ as follows.
The energy equation with radiative cooling is given as
\begin{equation}
    \frac{d}{dt}(3kn_{e}T)=-n_e^2\Lambda, \label{energy}
\end{equation}
where $n_e$, $T$, $k$,and $\Lambda$ are electron density, temperature, Boltzmann constant, and radiative cooling function, respectively.
By integrating Equation \ref{energy} from $T_0=10^4$ K to $T_1=10^7$ K with constant electron density, we obtain;

\begin{equation}
    \tau_\mathrm{rad}=\int_{T_0}^{T_1}3k/(n_e\Lambda) \mathrm{d}T=21.1\Big(\frac{n_e}{10^{11}~\mathrm{cm^{-3}}}\Big)^{-1}~\mathrm{minutes}.\label{Radeq}
\end{equation}
We calculated $\Lambda$ using CHIANTI 10.1 \citep{Dere1997A&AS..125..149D,Dere2023ApJS..268...52D} under the photospheric abundance.
Also, we note that the initial temperature is fixed at $10^7$ K. The temperatures obtained from GOES observation have only a factor difference (Section \ref{Re:params}, Table \ref{Elist}, Figure \ref{hist}), so this assumption is reasonable to simplify the discussion.
We relate $\tau_\mathrm{rad}$ and $EM$ through their dependence on electron density.
The emission measure $EM$ is written as 
\begin{equation}
    EM=\bar{n}_e^2L^3,    \label{EMeq}
\end{equation}
where $\bar{n}_e$ and $L$ are the spatially averaged electron density and the flare spatial scale, respectively.
As shown in H$\alpha$ images, H$\alpha$ postflare loops can be observed within only a localized region (Figures \ref{18cm01}, \ref{smart01}, \ref{FISCH01}, \ref{sddi01}).
This indicates that the electron density is distributed non-uniformly.
In regions with enhanced electron density, the radiative cooling timescale should be shorter than in other regions.
Therefore, the electron density in the areas where H$\alpha$ postflare loops are first observed is higher than the average density $\bar{n}_e$, and this enhanced electron density should be used in Equation \ref{Radeq}.
To consider this locally enhanced electron density, we relate the $n_e$ in Equation \ref{energy} and $\bar{n}_e$ as $n_e=\alpha\times\bar{n}_e$, where $\alpha$ is larger than unity but may not be so large. 
In \citet{Jejcic2018ApJ...867..134J}, they obtained electron density of postflare loops for the X8.2 flare on 2017 September 10, in which the electron density near the loop top is a few times larger than regions near the footpoints.
Similar results were also reported by \citet{Guidoni2015ApJ...800...54G}, who investigated an M1.3 flare on 2011 January 28. In their study, the electron density at the loop top was found to be 2–3 times larger than that at the loop legs (see Figure 10 in \citet{Guidoni2015ApJ...800...54G}).
Furthermore, \citet{Brose2022A&A...663A..18B} analyzed an M5.6 flare on 2015 January 13 and reported that the column emission measure at the loop top is about one order of magnitude larger than the loop-averaged column emission measure (see Figure 3 and Figure 9 in \citet{Brose2022A&A...663A..18B}). Since the column emission measure is proportional to the square of the electron density, a one-order difference in column emission measure corresponds to about a factor of three difference in electron density.
Although $\alpha$ may vary from flare to flare, we adopt a fixed value $\alpha = 3$ to simplify the analysis according to the observational results reported in the above references.
While the choice of $\alpha$ slightly changes the coefficients in the following equations, it does not significantly affect our conclusions.
By relating Equation \ref{Radeq} and Equation \ref{EMeq} using $n_e=3\times\bar{n}_e$, we obtained the equation for $\tau_\mathrm{rad}$ as a function of $EM$ and $L$;

\begin{equation}\label{tauEM}
    \tau_\mathrm{rad}=7.0\Big(\frac{EM}{10^{49}~\mathrm{cm}^{-3}}\Big)^{-1/2}\Big(\frac{L}{10~\mathrm{Mm}}\Big)^{3/2}~\mathrm{minutes}.
\end{equation}

As for the relation between $F_\mathrm{X}$ and $\tau_\mathrm{rad}$, we used the relation between $F_\mathrm{X}$ and $EM$ obtained in Appendix \ref{appendix:EMtoFx} under the assumption of $F_\mathrm{X}\propto EM$. The obtained $F_\mathrm{X}$-$\tau_\mathrm{rad}$ relation is
\begin{equation}\label{tauFx}
    \tau_\mathrm{rad}=7.0\Big(\frac{F_\mathrm{X}}{10^{-5}~\mathrm{W}~\mathrm{m}^{-2}}\Big)^{-1/2}\Big(\frac{L}{10~\mathrm{Mm}}\Big)^{3/2}~\mathrm{minutes}.
\end{equation}

In Figure \ref{AreaColor} (a) and (b), we plotted the relations $F_\mathrm{X}$-$\tau_\mathrm{rad}$ (Equation \ref{tauFx}) and $EM$-$\tau_\mathrm{rad}$ (Equation \ref{tauEM}), respectively.
The spatial scales are set as $L=50, 20, 7$ Mm.
These scaling relations generally agree with the observed correlation between $F_X$ ($EM$) and $\Delta t$, suggesting that radiative cooling can explain the timing of H$\alpha$ loops appearance with respect to the soft X-ray peak time.
Assuming $\tau_\mathrm{rad}\approx\Delta t$, we can estimate $n_e$ from $\Delta t$ using Equation \ref{Radeq}. The estimated $n_e$ ranges from $10^{10.8}$ to $10^{12.3}$ cm$^{-3}$, with a mean value of $10^{11.7}$ cm$^{-3}$.
These values are consistent with the electron densities obtained in previous studies (e.g., $10^{10.3}$ cm$^{-3}$ in \citet{Schmieder1995SoPh..156..337S}; $10^{11}$ cm$^{-3}$ in \citet{Heinzel1992SoPh..139...81H}; $10^{12}$ cm$^{-3}$ in \citet{Schmieder1987ApJ...317..956S};
up to $10^{13}$ cm$^{-3}$ in \citet{Jejcic2018ApJ...867..134J}).

After the GOES peaks, the electron density had already increased significantly due to the filling of loops with evaporated plasma, enhancing radiative cooling.
In \cite{Aschwanden2001SoPh..204...91A}, cooling of the X5.7 flare on 2000 July 14 (Bastille Day event) was investigated. During the decay phase after the GOES peak, radiative cooling was efficient, which is consistent with our statistical result.
On the other hand, in the early phase of the flare, before the GOES peak, conductive cooling was efficient mainly due to high temperature ($\ge30$ MK) \citep{Aschwanden2001SoPh..204...91A}. Although conductive cooling may also be important before the GOES peaks even in our flares, our result means that radiative cooling is the dominant mechanism determining the times from GOES peaks ($\sim10$ MK) to appearance of H$\alpha$ postflare loops ($\sim10^4$ K).
After the GOES peak, draining of materials along the loops can also remove energy as enthalpy flux \citep{Serio1991A&A...241..197S}.
However, our results suggest that radiative cooling is dominant and drains may have only minor effects in the cooling of loops.
In one-dimensional fluid simulations of cooling of loops, loops with a uniform cross-sectional area are strongly affected by draining, whereas loops with an expanding area are much less influenced by it \citep{Reep2022ApJ...933..106R,Reep2024ApJ...967...53R}.
Thus, our results are consistent with the idea of the expanding area of loops expected from decreasing magnetic field with the height \citep{Gary2001SoPh..203...71G}.

\subsection{Dependence on the spatial scale: method to estimate stellar flare size}
\begin{figure}[htbp]
\centering
\includegraphics[width=18cm]{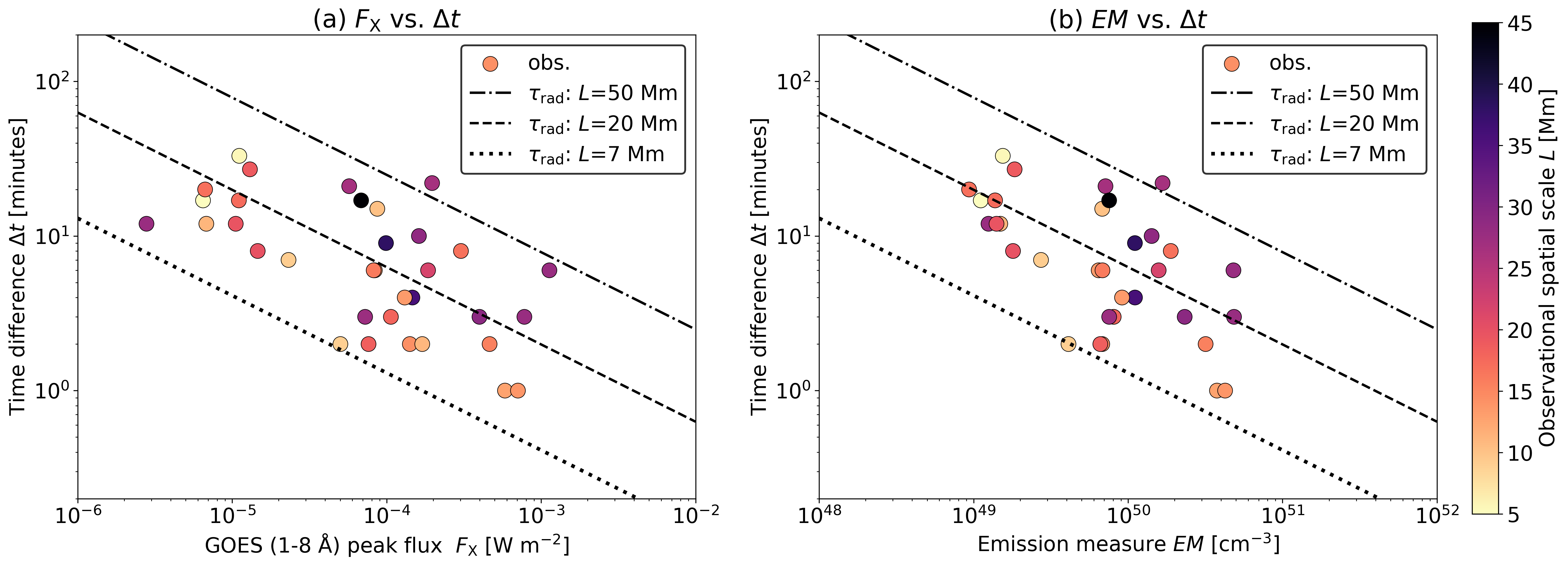}
\caption{The same as the Figure \ref{types}, but colored with the observed spatial scale. The black dashed-dotted, dashed, and dotted lines in (a) indicate the radiative cooling time ($\tau_\mathrm{rad}$) as a function of $F_\mathrm{X}$ for the spatial scales of 50 Mm, 20 Mm, and 7 Mm, respectively. The $\tau_\mathrm{rad}$ as a function of $EM$ is shown in (b) as the same format with (a). The range of color is set 5-45 Mm for visibility, although the range of the observed spatial scale is 5.1-50.4 Mm.}
\label{AreaColor}
\end{figure}

\begin{figure}[htbp]
\centering
\includegraphics[width=11.5cm]{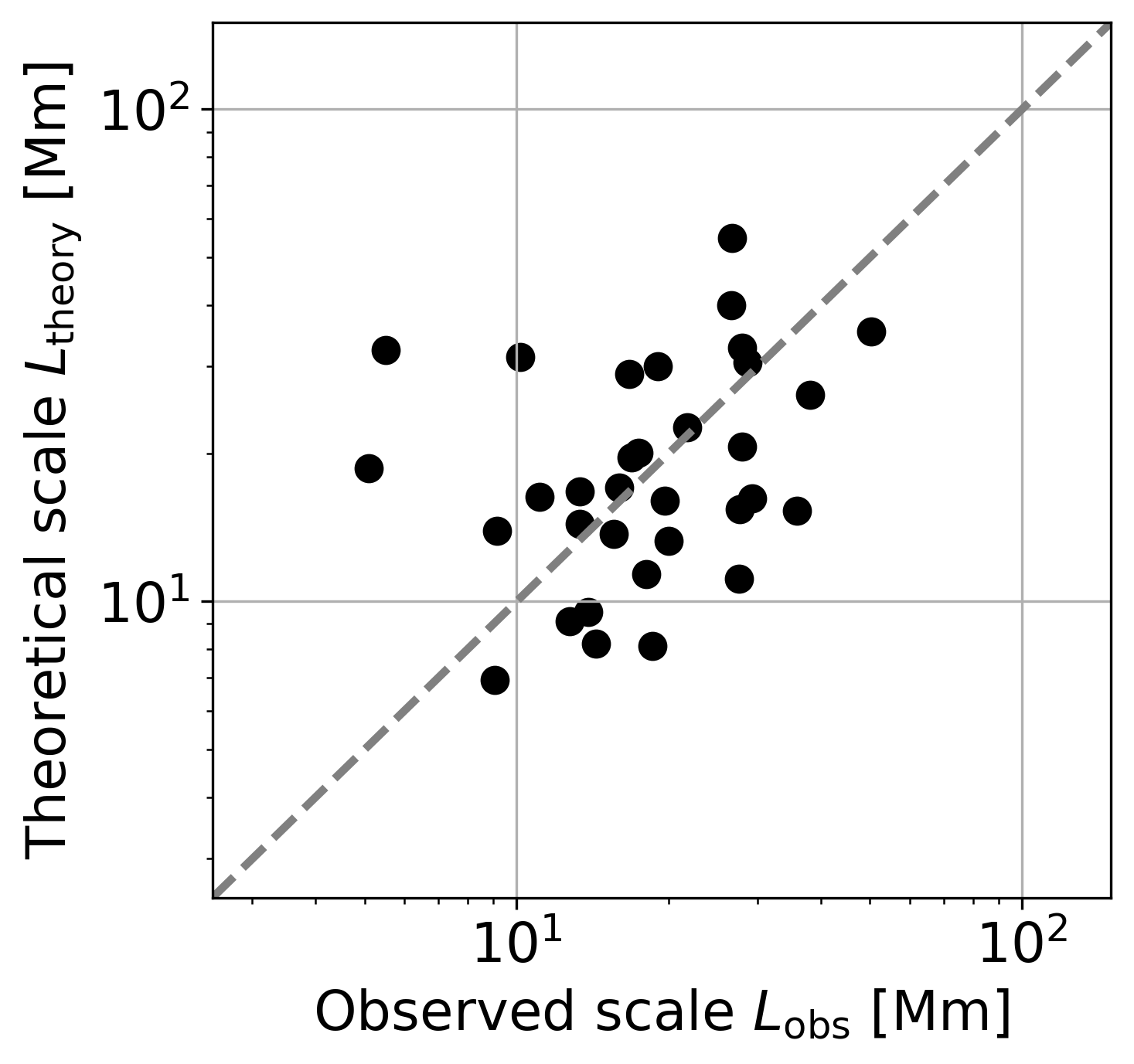}
\caption{The relationship between observed spatial scale and theoretical spatial scale. The gray dashed line indicates $y=x$ relation.}
\label{spatial}
\end{figure}

In the previous section, we compared $\Delta t$ and $\tau_\mathrm{rad}$ calculated using the observed $L$ and $EM$.
Consequently, our statistical results suggest that radiative cooling can generally explain the timing of the H$\alpha$ loops appearance with respect to the soft X-ray peak time.
Furthermore, in Figure \ref{AreaColor} (a) and (b), the theoretical lines of $\tau_\mathrm{rad}$ with larger $L$ lie above those with smaller $L$. This is consistent with the trend of the dependence of $F_\mathrm{X}-\Delta t$ and $EM-\Delta t$ relations on $L$: events with smaller (larger) $L$ tend to have shorter (longer) $\Delta t$ for a given $F_\mathrm{X}$ or $EM$.
This means that the theoretical relations in Equations \ref{tauFx} and \ref{tauEM} can account for the dependence of the $F_\mathrm{X}$–$\Delta t$ and $EM$–$\Delta t$ relationships on the spatial scale $L$.
In this section, to further examine this spatial scale dependence, we compare the observed spatial scale based on the imaging observations (hereafter $L_\mathrm{obs}$) with theoretical spatial scale suggested from observed $EM$ and $\Delta t$ using Equations \ref{tauEM} and \ref{tauFx} (hereafter $L_\mathrm{theory}$).
Using Equation \ref{tauEM} and assuming $\tau_\mathrm{rad}\approx\Delta t$, we can obtain $L_\mathrm{theory}$ as a function of $EM$ and $\Delta t$:
 \begin{equation}\label{Lth}
    L_\mathrm{theory}=10\Big(\frac{EM}{10^{49}~\mathrm{cm}^{-3}}\Big)^{1/3}\Big(\frac{\Delta t}{7.0~\mathrm{minutes}}\Big)^{2/3}~\mathrm{Mm}.
\end{equation}
From Equation \ref{tauFx}, $L_\mathrm{theory}$ can also be expressed as a function of $F_\mathrm{X}$ and $\Delta t$:
\begin{equation}
    L_\mathrm{theory}=10\Big(\frac{F_\mathrm{X}}{10^{-5}~\mathrm{W}~\mathrm{m}^{-2}}\Big)^{1/3}\Big(\frac{\Delta t}{7.0~\mathrm{minutes}}\Big)^{2/3}~\mathrm{Mm}.
\end{equation}
Figure \ref{spatial} shows the relation between $L_\mathrm{obs}$ (Table \ref{Elist}) and $L_\mathrm{theory}$. The correlation coefficient is 0.35.
As the overall trend, this relation shows $L_\mathrm{theory}\approx L_\mathrm{obs}$, although a few outliers with small $L_\mathrm{obs}$ values are present: particularly the two points with $L_\mathrm{obs}$ smaller than 6 Mm.
These two points correspond to off-limb events (events 4 and 24), and their three-dimensional structures were not taken into account in the estimation of the spatial scale, which possibly resulted in an underestimation of $L_\mathrm{obs}$.
When these two events are excluded, the correlation coefficient improves to 0.44.
To further evaluate this result, we calculated the root mean square of the difference between $L_\mathrm{obs}$ and $L_\mathrm{theory}$ (RMSD) : \
\begin{equation}
\mathrm{RMSD}=\sqrt{\frac{1}{N}\sum_{i=1}^{N}(L_{\mathrm{obs},i}-L_{\mathrm{theory},i})^2},
\end{equation}
where $N$ is the number of data and $i$ is the index of summation.
The RMSD is calculated to be 11.6 Mm for all data (10.6 Mm when the two outliers, event 4 and 24, are excluded).
This indicates that the Equation \ref{Lth} can sufficiently distinguish solar flares with small ($\sim10$ Mm) and large ($\sim100$ Mm) spatial scales.
In Figure \ref{AreaColor}, the dependence of $\Delta t$ on $L_\mathrm{obs}$ is weak around $F_\mathrm{X}\approx10^{-5}$ W m$^{-2}$ ($EM\approx10^{49}$ cm$^{-3}$).
The small value of $F_\mathrm{X}$ corresponds to a low electron density for a given spatial scale.
As a result, radiative cooling is potentially less effective, and other cooling mechanism such as conductive cooling could become relatively effective, for data around $F_\mathrm{X}\approx10^{-5}$ W m$^{-2}$. Such a suppression of radiative cooling may explain the weak dependence of $\Delta t$ on $L_\mathrm{obs}$ around $F_\mathrm{X}\approx10^{-5}$ W m$^{-2}$.

In this study, we classified some events as type ``u'', in which H$\alpha$ center images do not show bright postflare loops clearly but exhibit dark downflows in the red wing images.
On the other hand, for ``c'' events, downflows in the red wing images were observed as either bright or dark features.
Although ``c'' events commonly exhibited downflows at the almost same time as the appearance of bright loops in the H$\alpha$ center images, some ``c'' events showed downflows in the red wing images after bright loops appeared in the H$\alpha$ center images.
We note that there are no cases in which redshifted downflows appear earlier than the bright loops in the H$\alpha$ center images.
In our data set, the delay in the onset of downflows relative to the appearance of the bright loops in the H$\alpha$ center images is up to about 4 minutes.
Therefore, the measured $\Delta t$ of ``u'' type events based on dark downflows may have a systematic offset compared with that of type ``c'' events based on the bright loops. Since the theoretical spatial scale is expressed as $L_\mathrm{theory}\propto\Delta t^{2/3}$ (Equation \ref{Lth}), an offset in spatial scale ($\delta L_\mathrm{theory}$) due to an offset in time difference ($\delta(\Delta t)$) can be written as $\delta L_\mathrm{theroy}/L_\mathrm{theory}\approx\frac{2}{3}\times\delta(\Delta t)/\Delta t$. 
Here, we consider the ``u'' event with the typical time difference of $\Delta t=20$ minutes and estimated spatial scale of $L_\mathrm{theory}=20$ Mm.
If time difference $\Delta t$ have the offset of $\delta(\Delta t)= 4$ minutes, then $\delta L_\mathrm{theory}\approx2.7$ Mm. Such an offset in spatial scale is minor, and the effect of this potential discrepancy between type ``c'' and ``u'' events does not significantly affect our overall results.
Nevertheless, the time difference between the appearance of bright loops and dark downflows is crucial for further refinement of our scaling relations. Therefore, a more detailed investigation of this issue should be conducted in future studies.

As the required parameters to estimate $L_\mathrm{theory}$ using Equation \ref{Lth}, $EM$, or $F_\mathrm{X}$ as we discuss in Appendix \ref{appendix:EMtoFx} (Equation \ref{F:EM}), can be obtained even from stellar observations. In addition, the radiative cooling $\Delta t$ can be roughly estimated from the time difference between soft X-ray peak times and H$\alpha$ peak times corresponding to the appearance timing of postflare loops \citep{Otsu2024ApJ...974L..13O}.
Thus, these results mean that the spatial scale of flares can be estimated even for the unresolved stellar cases by using two quantities $F_\mathrm{X}$ (or $EM$) and $\Delta t$, which can be obtained from spatially integrated data.
The comparisons of $L_\mathrm{theory}$ with the ribbon distance $d$ and the square root of the ribbon area $A^{1/2}$ are also provided in Appendix \ref{appendix:ribbon} for reference.

Stellar flares can exhibit much larger energies than solar flares, and the electron density of stellar postflare loops is expected to reach on the order of $10^{12}-10^{13}$ cm$^{-3}$ or higher \citep{Heinzel2018ApJ...859..143H}.
In the case of such dense loops, radiative cooling can dominate the cooling process, and our scaling relations are expected to be successfully applicable in this regime. However, because stellar flares cannot be spatially resolved, the validity of our method cannot be directly confirmed for stellar cases in contrast to solar cases.
Therefore, it is critical to compare spatial scales of stellar fares obtained using our method with those derived from other approaches proposed in previous studies \citep{Shibata1999ApJ...526L..49S,Shibata2002ApJ...577..422S,Namekata2017ApJ...851...91N}.
In \cite{Shibata1999ApJ...526L..49S,Shibata2002ApJ...577..422S}, they explained the emission measure ($EM$)-temperature ($T$) relation of solar and stellar flares using a theory based on a magnetic reconnection model with heat conduction and chromospheric evaporation under the assumption of the gas pressure of a flare loop comparable to the magnetic pressure.
From their theoretical relationship, the spatial scale can be estimated from $EM$ and $T$.
The validation of this theory is further investigated using data from solar flares and small flares in solar quiet regions \citep{Namekata2017PASJ...69....7N,Kotani2023MNRAS.522.4148K}.
On the other hand, in \citet{Namekata2017ApJ...851...91N}, they explained flare energy ($E$)-flare duration ($\tau$) relation of solar and stellar white light flares using the magnetic reconnection theory. The $E$-$\tau$ relation can be used for the estimation of spatial scale of flares. In the present study, we focused on the cooling of flare loops from $\sim10^7$ to $10^4$ K and proposed a new method for estimating flare scales using $\tau_\mathrm{rad}$-$F_\mathrm{X}$ or $\tau_\mathrm{rad}$-$EM$ relation.
In future comparisons between solar and stellar flares,
it would be important to refine each physical theory by confirming the consistency of spatial scales estimated using various methods based on different physical perspectives.

\subsection{Toward statistical analysis of Sun-as-a-star H$\alpha$ light curves and spectra}\label{sub:ToSaS}

In this study, we used the GOES peak time as the typical timing of the appearance of evolved soft X-ray loops, while imaging data were used for determining the timing of H$\alpha$ loops appearance to secure sufficient data samples.
Thus, soft X-ray loops are traced in a spatially averaged manner, whereas H$\alpha$ loops are spatially resolved, and the successive formation of H$\alpha$ loops is not taken into account. 
To trace H$\alpha$ loops in a spatially averaged manner of successively formed loops, the Sun-as-a-star light curves of H$\alpha$ should be investigated.
In other words, the peak timing of the light curve corresponding to H$\alpha$ postflare loops should be used as $t_\mathrm{H\alpha}$.
As an example of Sun-as-a-star H$\alpha$ data, in \citet{Otsu2024ApJ...974L..13O}, the Sun-as-a-star analysis of the X1.6 flare on 2023 August 5 (event 23 in this study) was carried out. The time difference between a GOES soft X-ray peak time and a H$\alpha$ secondary peak due to postflare loops are obtained as approximately 13 minutes. On the other hand, the time difference between the GOES peak time and the appearance timing of postflare loops in H$\alpha$ image (i.e., $\Delta t$ in this study) for this event is about 10 minutes (Table \ref{Elist}). Thus, in the case of this event, it took about 3 minutes from the appearance of postflare loops to the peak of the light curve in H$\alpha$.
This time gap is likely caused by the successive formation of flare loops.
Although this gap is small compared to $\Delta t=10$ minutes, such a gap could cause systematic differences between the result in this study and results based on Sun-as-a-star H$\alpha$ data.
A statistical study of Sun-as-a-star H$\alpha$ data exhibiting secondary peaks due to postflare loops should be performed to examine the effect of successive flare loop formations.
We also note the importance of Doppler signatures coming from downflows along postflare loops. Even when a secondary peak cannot be seen in an H$\alpha$ light curve, Doppler shifted components from downflows may be identified in H$\alpha$ spectra as in Event 4 in \citet{OtsuETAL2022} (event 13 in this study). Thus, the appearance timing of downflow signatures can also be used to discuss the cooling of loops from soft X-ray temperatures.
In addition, downflows can represent specific motions of plasma along the postflare loops \citep{Wollmann2023A&A...669A.118W,Otsu2024ApJ...974L..13O}. Therefore, simultaneously detecting downflows and confirming the consistency between their appearance and radiative cooling would provide strong evidence for the dominant role of postflare loops in stellar flares.

By quantitatively analyzing the timing of the secondary peak in the stellar H$\alpha$ light curve based on the scaling law proposed in this paper, stellar postflare loops can be clearly detected.
Furthermore, the accurate detection of stellar postflare loops is also helpful for the study of stellar filament/prominence eruptions and CMEs.
Since stellar CMEs are thought to affect the environments of exoplanets, their detection has been extensively investigated through collaborations between the stellar and exoplanetary communities.
Some Doppler-shifted components in H$\alpha$ spectra associated with stellar flares were likely caused by stellar filament or prominence eruptions, which can serve as indirect evidence of stellar CMEs.
However, downflows along postflare loops can also cause Doppler-shifted components, making it difficult to identify whether stellar filament/prominence eruptions have occurred, particularly in low-velocity cases.
If the detection of stellar postflare loops is clarified for the stellar flares with Doppler-shifted components by confirming that the timing of the secondary peak in stellar H$\alpha$ data is consistent with the radiative cooling timescale, the possibility of stellar filament or prominence eruptions can be ruled out.
In contrast, if the timing of the secondary peak in stellar H$\alpha$ data clearly differs from the radiative cooling timescale, the Doppler-shifted components are likely caused by filament or prominence eruptions, even in low-velocity cases.
In this way, our study offers important implications not only for stellar postflare loops but also for the understanding of stellar filament or prominence eruptions and CMEs.

\section{Conclusions}\label{Sec:conc}
In this study, we statistically investigated the appearance timings of H$\alpha$ loops ($t_\mathrm{H\alpha}$) with respect to the soft X-ray peak times ($t_\mathrm{X}$).
As a result, we found a negative correlation between the soft X-ray peak flux $F_\mathrm{X}$ (and the corresponding emission measure, $EM$) and the time difference $\Delta t = t_\mathrm{H\alpha} - t_\mathrm{X}$.
The $F_\mathrm{X}$–$\Delta t$ and $EM$–$\Delta t$ relations are consistent with the scaling law based on the radiative cooling time scale $\tau_\mathrm{rad}$.
Recent temporally resolved stellar observations have revealed secondary peaks in the H$\alpha$ light curves of stellar flares \citep[e.g.,][]{Ichihara2025PASJ..tmp...85I}.
To quantitatively investigate the origin of such secondary peaks in stellar H$\alpha$ light curves, the scaling relation based on radiative cooling derived in the present study is particularly useful for determining whether stellar postflare loops are the dominant source of radiation.
We also proposed a method to estimate the spatial scale of flares using the derived scaling law. 
Our approach enables the spatial scale of unresolved stellar flares to be estimated from $F_\mathrm{X}$ (or $EM$) and $\Delta t$. In future applications to stellar flare studies, it will be essential to confirm the consistency of the spatial scale obtained through multiple independent methods based on different physical perspectives.
Since stellar flares cannot be spatially resolved, the uncertainties of each method may not be negligible. However, if the spatial scales estimated by multiple independent methods are consistent, the reliability of the estimated results will increase.

\begin{acknowledgments}
We express our sincere gratitude to the staff of Hida and Kwasan Observatories for developing and maintaining the instruments and daily observation. 
We would like to acknowledge the data use from GOES. 
This work was supported by JSPS KAKENHI grant Nos. JP24K07093  (PI: A. A.), JP24K00680 (PI: K. S.), JP24K17082 (PI: K. I.), and JP24H00248 (PI: D. Nogami).
This work was also supported by JST SPRING, Grant Number JPMJSP2110 (T.O.), and by a grant from the Hayakawa Satio Fund awarded by the Astronomical Society of Japan.
\end{acknowledgments}

\software{sunpy \citep{Sunpy2020ApJ}, ChiantiPy \citep{Dere2013ascl.soft08017D}}

\clearpage
\begin{table}[htbp]
\centering
\caption{Flare List}

\begin{tabular}{ccccccccccccc}
\hline
\#& 
$t_{\mathrm{X}}$\footnote[1]{Peak time of GOES/XRS-B (1-8 {\AA}).}&
$t_\mathrm{H\alpha}$\footnote[2]{Appearance time of postflare loops in H$\alpha$ images.}&
$\Delta t$\footnote[3]{Time difference between GOES peak time and appearance time of H$\alpha$ postflare loops ($\Delta t=t_\mathrm{H\alpha}-t_\mathrm{X}$).}&
$F_\mathrm{X}$\footnote[4]{GOES/XRS-B flux at GOES peak time. The recalibrated values are used (see the text for details). The corresponding GOES classes are given in parentheses.}&
$EM$\footnote[5]{Volume emission measure calculated from GOES/XRS-A and -B data at GOES peak time.}&
$T$\footnote[6]{Temperature calculated from GOES/XRS-A and -B data at GOES peak time.}&
$d$\footnote[7]{Ribbon distance. See the text for details.}&
$A^{1/2}$\footnote[8]{Square root of ribbon area. See the text for details.}&
$L$\footnote[9]{Spatial scale. See the text for details; $L=((\pi d/2)\times(A/2))^{1/3}$. For off-limb events, the square root of the loops area are used.} &
H$\alpha$ obs.\footnote[10]{The instrument for H$\alpha$ imaging data. 'Sart.' stands for 'Sartrius'.}&
type\footnote[11]{Type of each event. The events with ``c'' showed bright postflare loops in H$\alpha$ center image clearly. For the events with ``u'', postflare loops are unclear in H$\alpha$ center images.
The events with ``off'' showed H$\alpha$ postflare loops at the off-limb regions.}&
NOAA\\
&(UT) & (UT)& (min) &(W m$^{-2}$) &($10^{50}$ cm$^{-3}$) & ($10^7 $K)&(Mm)&(Mm)&(Mm) & & &\\

\hline
\hline

1& 2004-11-06 00:34& 00:36& 2  &$1.4\times10^{-4}$ (X1.4)& 0.68 & 2.28 & 22.3 & 13.0 & 14.4 &Sart. & c &10696 \\
2& 2004-11-10 02:13& 02:16& 3  &$4.0\times10^{-4}$ (X4.0)& 2.33 & 1.90 & 17.0 &43.2 &29.2 &Sart. & c &10696\\
3& 2005-01-20 07:00& 07:06& 6  &$1.1\times10^{-3}$ (X11)& 4.8 & 2.58 & 32.0 &29.4 &27.9 &Sart. & c &10720\\
4& 2010-08-18 05:45& 06:02& 17  &$6.5\times10^{-6}$ (C6.5)& 0.11 & 0.85 & $\cdots$ &$\cdots$ &5.1 &T1 & off &$\cdots$\\
5& 2011-09-06 22:20& 22:28& 8  &$3.0\times10^{-4}$ (X3.0)& 1.89 & 1.76 & 7.9 &27.5 &16.7 &T1 & c &11283\\
6& 2012-03-07 00:24& 00:27& 3  &$7.8\times10^{-4}$ (X7.8)& 4.85 & 1.77 & 35.1 &28.2 &28.0 &T1 & c &11429\\
7& 2012-03-07 01:15& 01:37& 22  &$2.0\times10^{-4}$ (X2.0)& 1.67 & 1.33 & 36.8 &25.6 &26.7 &T1 & c &11429\\
8& 2013-03-21 22:04& 22:11& 7  &$2.3\times10^{-5}$ (M2.3)& 0.27 & 1.05 & $\cdots$ &$\cdots$ &9.2 &T1 & off &$\cdots$\\
9& 2013-05-14 01:11& 01:13& 2  &$4.6\times10^{-4}$ (X4.6)& 3.18 & 1.61 & 33.8  &11.9 &15.6 &FISCH & c &11748\\
10& 2013-05-15 01:48& 01:54& 6  &$1.9\times10^{-4}$ (X1.9)& 1.58 & 1.33 & 21.4 &24.7 &21.7 &FISCH & c &11748\\
11& 2013-10-28 02:03& 02:07& 4  &$1.5\times10^{-4}$ (X1.5)& 1.11 & 1.47 & 33.5 &41.8 &35.8 &T1 & c &11875\\
12& 2017-04-02 08:02& 08:08& 6  &$8.4\times10^{-5}$ (M8.4)& 0.65 & 1.43 & 16.1 &13.7 &13.3 &SDDI & c &12644\\
13& 2021-04-19 23:42& 23:59& 17  &$1.1\times10^{-5}$ (M1.1)& 0.14 & 1.01 & 23.3 &17.0 &17.4 &SDDI & u &12816\\
14& 2022-03-25 05:26& 05:34& 8 &$1.5\times10^{-5}$ (M1.5)& 0.18 & 1.02 & 38.3 &16.3 &20.0 &SDDI & u &12974\\
15& 2022-10-02 02:21& 02:36& 15 &$8.7\times10^{-5}$ (M8.7)& 0.68 & 1.42 & 8.1 &12.9 &10.2 &SDDI & u &13110\\
16& 2023-01-11 06:09& 06:36& 27  &$1.3\times10^{-5}$ (M1.3)& 0.18 & 0.94 & 38.8 &15.0 &19.0 &SDDI & u &13181\\
17& 2023-04-01 03:28& 03:40& 12  &$6.8\times10^{-6}$ (C6.8)& 0.15 & 0.75 & $\cdots$ &$\cdots$ &11.1 &SDDI & off &$\cdots$\\
18& 2023-04-28 01:22& 01:42& 20  &$6.7\times10^{-6}$ (C6.7)& 0.09 & 0.94 & 33.1 &13.6 &16.9 &SDDI & u &13285\\
19& 2023-05-09 00:33& 00:45& 12  &$2.8\times10^{-6}$ (C2.8)& 0.13 & 0.55 & 46.7 &24.0 &27.7 &SDDI & u &13299\\
20& 2023-05-09 20:52& 20:54& 2  &$5.0\times10^{-5}$ (M5.0)& 0.41 & 1.36 & 16.4 &7.6 &9.0 &SDDI & c &13296\\
21& 2023-07-02 23:14& 23:17& 3  &$1.1\times10^{-4}$ (X1.1)& 0.81 & 1.45 & 32.5 &15.2 &18.0 &SDDI & c &13354\\
22& 2023-07-18 00:06& 00:27& 21\footnote[12]{Events 22, 33, and 34 showed H$\alpha$ postflare loops before the GOES peak times, although they are typically faint. Therefore, the appearance times of the brighter groups of H$\alpha$ postflare loops are used.}  &$5.7\times10^{-5}$ (M5.7)& 0.71 & 1.01 & 59.7 &20.0 &26.6 &SDDI & u &13363\\
23& 2023-08-05 22:21& 22:31& 10  &$1.6\times10^{-4}$ (X1.6)& 1.42 & 1.29 & 27.3 &33.1 &28.6 &SDDI & c &13386\\
24& 2023-08-26 22:50& 23:23& 33  &$1.1\times10^{-5}$ (M1.1)& 1.56 & 0.95 & $\cdots$ &$\cdots$ &5.5 &SDDI & off &$\cdots$\\
25& 2024-01-29 04:38& 04:55& 17  &$6.8\times10^{-5}$ (M6.8)& 0.75 & 1.09 & 67.2 &49.2 &50.4 &SDDI & c &13559\\
26& 2024-02-14 03:10& 03:22& 12  &$1.1\times10^{-5}$ (M1.1)& 0.14 & 0.97 & 34.0 &16.8 &19.6 &SDDI & u &13582\\
27& 2024-05-03 02:22& 02:24& 2 &$1.7\times10^{-4}$ (X1.7)& $\cdots$\footnote[13]{soft X-ray data for event 27 are available only from GOES-18. However, {\it sunkit\_instruments} version 0.5.0 is not compatible with GOES-18 data. Therefore, event 27 was excluded from the calculation of $EM$ and $T$.} & $\cdots$$^{\mathrm{m}}$ & 9.8 &13.2 &11.0 &FISCH & c &13663\\
28& 2024-05-05 06:01& 06:05& 4  &$1.3\times10^{-4}$ (X1.3)& 0.91 & 1.56 & 16.4 &13.6 &13.3 &SDDI & c &13663\\
29& 2024-05-11 01:23& 01:24& 1  &$5.8\times10^{-4}$ (X5.8)& 3.76 & 1.68 & 13.1 &14.2 &12.7 &FISCH & c &13664\\
30& 2024-05-17 21:08& 21:11& 3  &$7.3\times10^{-5}$ (M7.3)& 0.75 & 1.14 & 46.8 &23.8 &27.5 &SDDI & u&13686\\
31& 2024-06-08 01:49& 01:58& 9 &$9.9\times10^{-5}$ (M9.9)& 1.11 & 1.07 & 47.5 &38.5 &38.1 &SDDI & c &13697\\
32& 2024-08-01 07:09& 07:15& 6  &$8.2\times10^{-5}$ (M8.2)& 0.68 & 1.35 & 19.4 &16.3 &16.0 &SDDI & c &13768\\
33& 2024-10-01 00:00& 00:02& 2$^{\mathrm{l}}$ &$7.6\times10^{-5}$ (M7.6)& 0.66 & 1.30 & 29.8 &16.5 &18.5 &SDDI & c &13842\\
34& 2024-10-01 22:20& 22:21& 1$^{\mathrm{l}}$  &$7.1\times10^{-4}$ (X7.1)& 4.25 & 1.80 & 16.4 &14.4 &13.9 &SDDI & c &13842\\
\hline
\end{tabular}

\label{Elist}
\end{table}
\clearpage


\clearpage

\appendix
\section{Ribbon distance  and Ribbon area}\label{appendix:ribbon}

\begin{figure}[htbp]
\centering
\includegraphics[width=12cm]{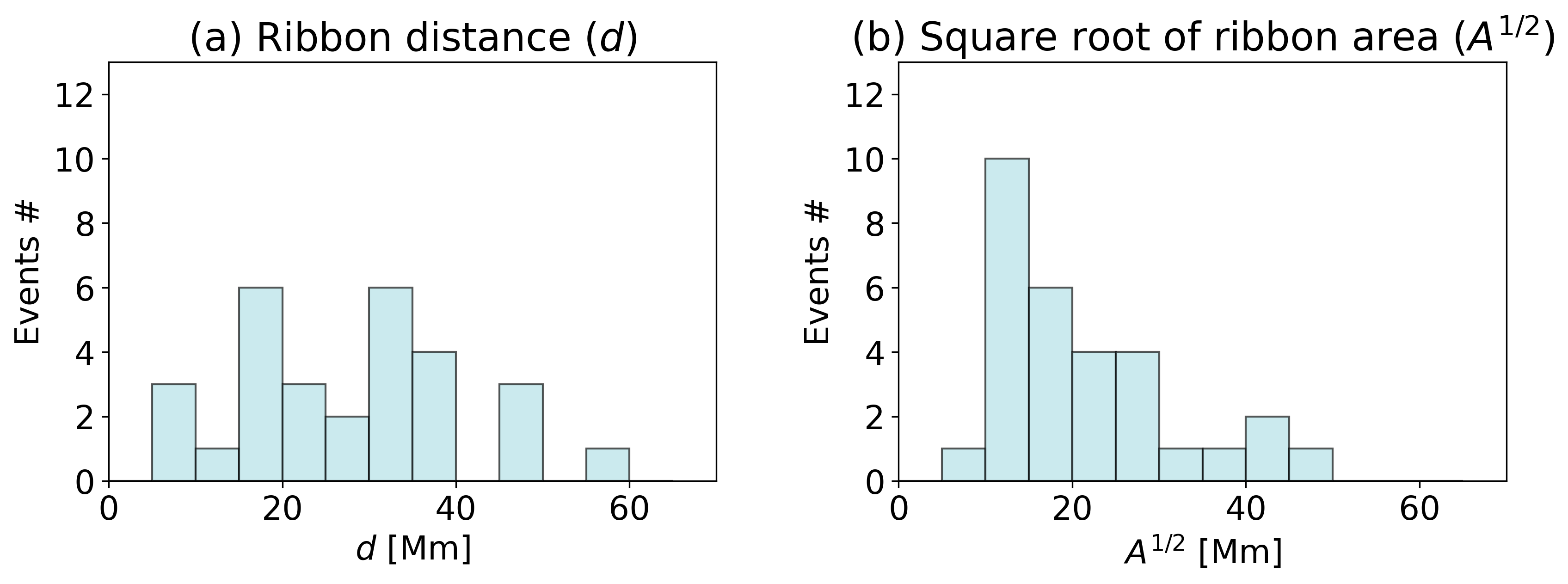}
\caption{Histograms of ribbon distance and square root of ribbon area. (a) Ribbon distance ($d$). (b) Square root of ribbon distance ($A^{1/2}$). We note that both panels include 30 events, excluding off-limb events (events 4, 8, 17, 24).}
\label{appendix:hist}
\end{figure}

\begin{figure}[htbp]
\centering
\includegraphics[width=12cm]{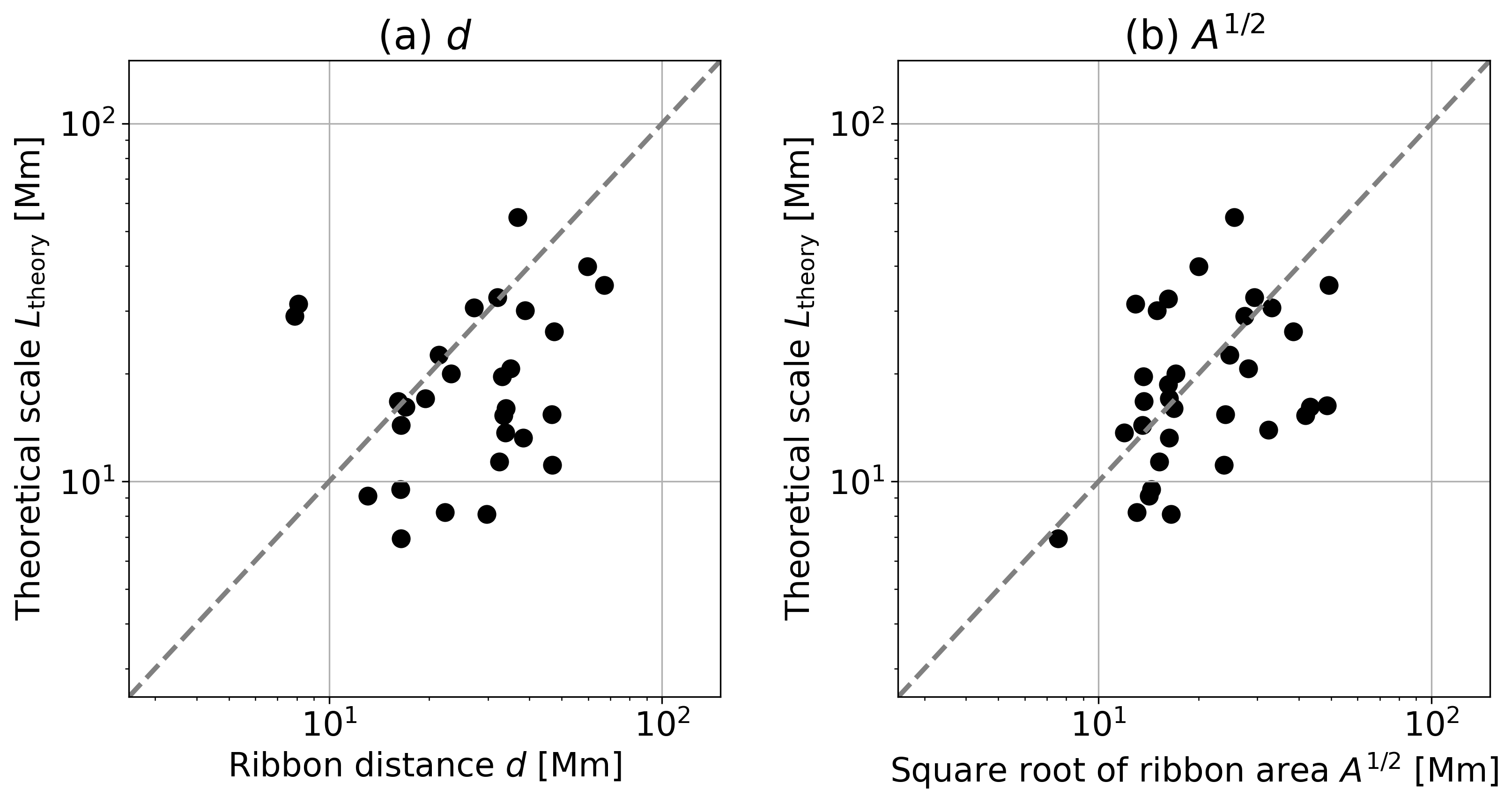}
\caption{The same as Figure \ref{spatial} but for ribbon distance ($d$) and square root of ribbon area ($A^{1/2}$). (a)The relation of $d$ and the theoretical spatial scale. (b)The relation of $A^{1/2}$ and theoretical spatial scale. We note that both panels include 29 events, excluding off-limb events (events 4, 8, 17, 24) and event 27.}
\label{appendix:scale}
\end{figure}

In this section, we present supplemental results for the ribbon distance $d$ and the square root of the ribbon area $A^{1/2}$, as derived in Section \ref{M:spatial}.
Figure \ref{appendix:hist} (a) and (b) show the distributions of $d$ and $A^{1/2}$, respectively.
The ribbon distance $d$ ranges from 7.9 to 67.2 Mm, with an average of approximately 29.2 Mm.
Both $d$ and $A^{1/2}$ are of the same order of magnitude as the spatial scale $L = \left[(\pi d/2) \times (A/2)\right]^{1/3}$.
Figure \ref{appendix:scale} (a) and (b) show the comparisons of the theoretical spatial scale $L_\mathrm{theory}$ with $d$ and $A^{1/2}$.
Both $d$ and $A^{1/2}$ roughly agree with $L_\mathrm{theory}$, as is the case with $L_\mathrm{obs}$.
We note that off-limb events are excluded in Figure \ref{appendix:scale}.
The correlation coefficients for $d$-$L_\mathrm{theory}$ and $A^{1/2}$-$L_\mathrm{theory}$ relations are $0.34$ and $0.39$, respectively. These results indicate that the choice of flare spatial scale ($d$, $A^{1/2}$, or $L_\mathrm{obs}$) does not significantly affect the comparison with theoretical expectations.

\clearpage
\section{GOES XRS-B flux ($F_X$) vs. emission measure ($EM$)}\label{appendix:EMtoFx}

\begin{figure}[htbp]
\centering
\includegraphics[width=9cm]{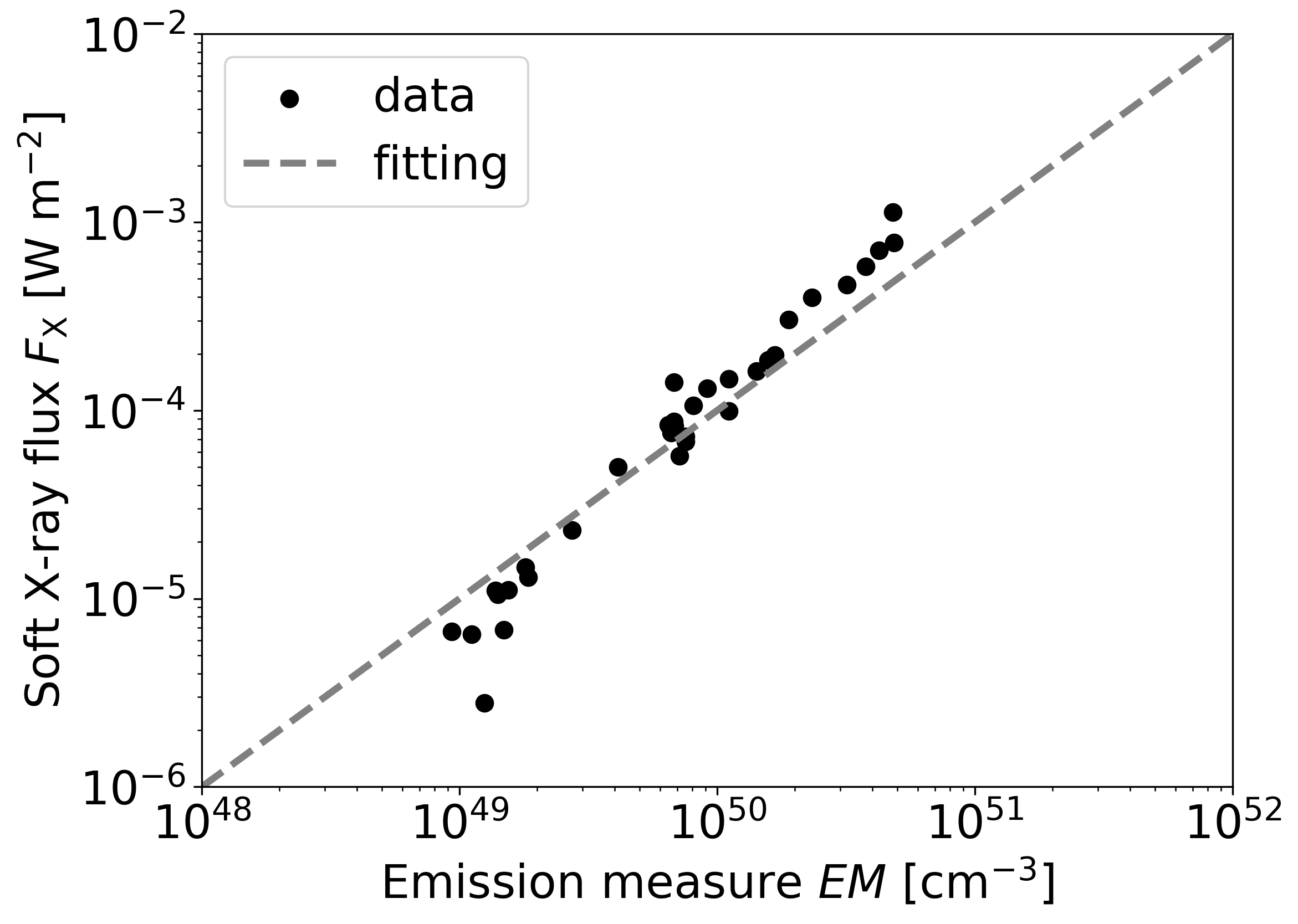}
\caption{The relation between emission measure calculated from GOES two bands, and GOES soft X-ray (1-8 {\AA}) flux. The data at the GOES peak time were used. The fitting result with equation \ref{F:EM} is shown with the gray dashed line. We note that both panels include 33 events, excluding event 27.}
\label{appendix:EMFx}
\end{figure}

Although $EM$ was originally obtained from the two bands of GOES/XRS (see Section \ref{M:GOES}), we derived a one-to-one relationship between $EM$ and $F_\mathrm{X}$ (GOES/XRS-B flux) for simply converting $F_\mathrm{X}$ to $EM$.
Figure \ref{appendix:EMFx} shows the relationship between $F_\mathrm{X}$ and $EM$ at the peak times of $F_\mathrm{X}$ for the events analyzed in this study. We use the relation $F_\mathrm{X}=10^b\times EM$ with a constant factor $b$, based on the assumption of a fixed temperature ($10^{7}$ K) (see Section \ref{Dis:scaling}).
By fitting the data in log scale with this relation ($\log_{10}(F_\mathrm{X}[\mathrm{W~m}^{-2}]) =\log_{10}{(EM[\mathrm{cm}^{-3}])}+b$), we obtained $b=-53.98\pm0.04\approx-54$.
The obtained relation can be rewritten as follows;

\begin{equation}\label{F:EM}
\displaystyle
    F_\mathrm{X} = 1\times10^{-5} (\frac{EM}{10^{49} \mathrm{cm}^{-3}})~[\mathrm{W~m}^{-2}],
\end{equation}
This relation is plotted as the gray dashed line in Figure \ref{appendix:EMFx}.
We used Equation \ref{F:EM} to convert $EM$ to $F_\mathrm{X}$ in Section \ref{Dis:scaling}.

\clearpage
\bibliography{sample701}{}
\bibliographystyle{aasjournalv7}

\end{document}